\documentclass[prb,aps,amssymb,twocolumn,superscriptaddress,10pt]{revtex4-2}
\usepackage{amsmath,amssymb,amsthm,bm}
\usepackage{amsfonts,xfrac}
\usepackage{xcolor,array}
\usepackage{hyperref}
\usepackage{graphicx}
\usepackage{subfigure}
\usepackage{tikz}
 
\newcommand{\Tr}{{\rm Tr}\,}

\renewcommand{\vec}[1]{\bm{#1}}
\newcolumntype{C}[1]{>{\centering\let\newline\\\arraybackslash\hspace{0pt}}m{#1}}

\begin{document}

\title{Universal signatures of Dirac fermions in entanglement and charge fluctuations}

\author{Valentin Cr\'epel}
\affiliation{Department of Physics, Massachusetts Institute of Technology, 77 Massachusetts Avenue, Cambridge, MA, USA
}

\author{Anna Hackenbroich}
\affiliation{Max-Planck-Institute of Quantum Optics, Hans-Kopfermann-Stra{\ss}e 1, 85748 Garching, Germany}
\affiliation{Munich Center for Quantum Science and Technology, Schellingstra{\ss}e 4, 80799 M{\"u}nchen, Germany}
	
\author{Nicolas Regnault}
\affiliation{Joseph Henry Laboratories and Department of Physics, Princeton University, Princeton, New Jersey 08544, USA}
\affiliation{Laboratoire de Physique de l'Ecole normale sup\'{e}rieure, ENS, Universit\'{e} PSL, CNRS, Sorbonne Universit\'{e}, Universit\'{e} Paris-Diderot, Sorbonne Paris Cit\'{e}, Paris, France}
	
\author{Benoit Estienne}
\affiliation{Sorbonne Universit\'e, CNRS, Laboratoire de Physique Th\'eorique et Hautes Energies, LPTHE, F-75005 Paris, France
}

\date{\today}

\begin{abstract}
We investigate the entanglement entropy (EE) and charge fluctuations in models where the low energy physics is governed by massless Dirac fermions. We focus on the response to flux insertion which, for the EE, is widely assumed to be universal, \emph{i.e.}, independent of the microscopic details.  We provide an analytical derivation of the EE and charge fluctuations for the seminal example of graphene, using the dimensional reduction of its tight-binding model to the one-dimensional Su-Schrieffer-Heeger model. Our asymptotic expression for the EE matches the conformal field theory prediction. We show that the charge variance has the same asymptotic behavior, up to a constant prefactor. To check the validity of universality arguments, we numerically consider several models, with different geometries and number of Dirac cones, and either for strictly two-dimensional models or for gapless surface mode of three-dimensional topological insulators. We also show that the flux response does not depend on the entangling surface geometry as long as it encloses the flux. 
Finally we consider the universal corner contributions to the EE. We show that in the presence of corners, the  Kitaev-Preskill subtraction scheme provides non-universal, geometry dependent results.
\end{abstract}

\maketitle

\section{Introduction}

Emergent Dirac fermions have become ubiquitous in modern condensed matter physics. Beyond the seminal case of graphene, massless Dirac fermions can be found in more exotic situations such as the surface of a three-dimensional topological insulator~\cite{RevModPhys.82.3045,RevModPhys.83.1057}, optical lattices~\cite{Tarruell_2012}, microwave experiments~\cite{PhysRevLett.110.033902} and even quasi-2D organic materials~\cite{doi:10.1143/JPSJ.75.054705,doi:10.1143/JPSJ.76.034711,PhysRevB.78.045415}. Dirac physics also appears in strongly correlated quantum systems such as quantum spin liquids~\cite{Savary_2016,Zhou-RevModPhys.89.025003} where evidence of gapless Dirac quantum spin liquids have been observed~\cite{Zhu2018EntanglementSO,Hu_2019}. By their inherent quantum many-body nature, the study of these systems heavily relies on numerical simulations. There, entanglement measurements such as the entanglement entropy (EE) and bipartite fluctuations have emerged as fundamental and powerful techniques to probe quantum phases. 

The success of EE and bipartite fluctuations is widespread. For one-dimensional systems, they can reliably detect quantum phase transitions, measure the central charge of critical points~\cite{HOLZHEY1994443,PhysRevLett.90.227902,calabrese2004entanglement} as well as the Luttinger parameter~\cite{PhysRevB.82.012405,PhysRevB.85.035409,Rachel_2012,Swingle_2013}. Furthermore in the vicinity of a quantum critical point, they provide a measure of the correlation length. For two-dimensional gapped systems, EE is capable of detecting intrinsic topological order and extracting the quantum dimension of the various anyonic excitations~\cite{PhysRevLett.96.110404,Levin_2006}. It can also identify the presence and nature of massless edge modes~\cite{PhysRevB.88.155314,Cr_pel_2019a,Cr_pel_2019b,PhysRevLett.123.126804,Estienne_2020}, and even massless hinge modes for three-dimensional insulators~\cite{hackenbroich2020fractional}. 

In this paper, we aim to find universal signatures of Dirac fermions in both the quantum EE and the bipartite charge fluctuations. A promising idea to detect Dirac matter is to use the entanglement response to flux insertions~\cite{metlitski2009entanglement,chen2017two,Zhu2018EntanglementSO}. This twist dependence of the EE however has been predicted using conformal field theory. While it is generally believed that this response is universal, that is insensitive to short-distance physics, a strong argument is still lacking. Conversely, there is \emph{a priori} no guarantee that the flux response is not going to be plagued by non-universal contributions in a given lattice model. In order to further support the claim that the flux response of the EE is robust, we investigate this response for various tight-binding models whose universal low-energy physics is described by Dirac fermions, such as graphene. For the latter, we provide an analytical derivation of the EE and its flux dependence from the tight-binding model and its relation to the one-dimensional Su-Schrieffer-Heeger (SSH) model. Furthermore for non-interacting fermions, the EE is tied to the statistics of charge fluctuations~\cite{Klich_2009a,Klich_2009b,Song_2011,PhysRevB.85.035409,Calabrese_2012}. Thus we propose and test an even simpler signature for Dirac fermions than the EE, namely the flux-dependence of the charge fluctuations.  

This article is organised as follows. In Sec.~\ref{sec:Methodology}, we provide a brief overview of the EE and particle fluctuations for non-interacting models. We also recall the exact results known for the one-dimensional SSH model and derive the analytical expression of the charge variance for this model. In Sec.~\ref{sec:GrapheneExact} we compute analytically the exact flux dependence of both the EE and the particle fluctuations for graphene. The strategy underlying this computation is that of dimensional reduction~\cite{PhysRevB.62.4191,Murciano_2020}, which allows to reduce the problem to a sum of one-dimensional SSH chains. We further argue that the flux dependence is in fact exact for any non-interacting tight-binding Hamiltonian in the same universality class. In Sec.~\ref{sec_extension_num} we benchmark our analytical predictions against numerical computations for several lattice models, including for a model of massless surface modes for a three-dimensional insulator. We also check the robustness of the flux response to deformations of the region considered. 
In Sec.~\ref{sec_KP} we analyse the effect of corners to the EE and the consequences for a potential Kitaev-Preskill subtraction scheme.

\section{Methodology} \label{sec:Methodology}

In this section, we provide an overview of the correlation matrix technique to compute the EE and charge fluctuations in non-interacting fermionic systems. We then discuss in detail the SSH model, including the asymptotic expression and finite size effects for both aforementioned quantities.

\subsection{Entropy and charge fluctuations in non-interacting fermionic systems} \label{ssec:PeschelTrick}

We consider a free fermionic lattice model with translation invariance, which is described by the generic Hamiltonian
\begin{equation}
\mathcal{H} = \sum_{\substack{\vec{r}, \vec{r}' \in \mathcal{L} \\ \tau, \tau' = 1 \cdots d}} c_\tau^\dagger (\vec{r}) h_{\tau \tau'} (\vec{r}-\vec{r}') c_{\tau'} (\vec{r}') \, ,
\end{equation}
where $c_\tau (\vec{r})$ denotes the fermionic annihilation operator for the state $\tau$ in the unit cell located at $\vec{r}$ in the lattice $\mathcal{L}$, and $d$ is the number of inequivalent quantum states within each unit cell. 
Exploiting translational symmetry, the Hamiltonian matrix $h(\vec{r})$ is conveniently expressed by its Fourier transform
\begin{equation}
\tilde{h} (\vec{k}) = \sum_{\vec{r}\in \mathcal{L}} h(\vec{r})  e^{-i \vec{k}\cdot \vec{r}} \, ,
\end{equation}
with $\vec{k} \in BZ$ in the first Brillouin zone.

At thermal equilibrium, the many-body system is described by a Gaussian density matrix $\rho_T = \exp \left( -\beta \mathcal{H} \right) / \mathcal{Z}$, with $\beta$ the inverse temperature and $\mathcal{Z}= \Tr(e^{-\beta \mathcal{H}})$. Note that in this article, we will always assume a zero temperature, meaning that $\rho_T$ becomes the projector onto the system's ground state. This Gaussian character is handed down to any subsystem of the original lattice. 
In other words, the reduced density matrix for a subregion $\mathcal{A}$
\begin{equation}
\rho_\mathcal{A} = \Tr_{\bar{\mathcal{A}}} (\rho_T) \, ,
\end{equation}
with $\Tr_{\bar{\mathcal{A}}} $ the partial trace over $\bar{\mathcal{A}}$ the complement of $\mathcal{A}$, is also Gaussian. 
As a consequence, Wick's theorem applies and all expectation values in $\mathcal{A}$ can be computed from the sole knowledge of the correlation matrix~\cite{wick1950evaluation}
\begin{equation} \label{eq:Method_DefinitionCorrelMatrix}
\left[ C_\mathcal{A} \right]_{\tau \tau'} (\vec{r}, \vec{r}') = \Tr_{\mathcal{A}} \left( \rho_\mathcal{A} \, c_\tau^\dagger (\vec{r}) c_{\tau'} (\vec{r}') \right) \, .
\end{equation}
Indeed, the relation~\cite{Chung_2001,Peschel_2003,cheong2004many,Peschel_2009,Peschel_2012}
\begin{equation} \label{eq:Method_densitymatrixcorrel}
\rho_\mathcal{A} = \det (1 - C_\mathcal{A}) \exp \left\{ c^\dagger \log \left[ C_\mathcal{A} \left(1-C_\mathcal{A}\right)^{-1} \right] c \right\},
\end{equation}
grants access to the entire eigen-decomposition of $\rho_\mathcal{A}$ from that of $C_\mathcal{A}$. Here, the summation over the omitted indices $\vec{r}$ and $\tau$ is assumed in the exponential. All observables of the \emph{many-body} problem can be evaluated from the diagonalization of the \emph{one-body} operator $C_\mathcal{A}$.

This expression is particularly useful when characterizing the properties of the free-fermion system, as it provides an efficient way to compute the EE of the region $\mathcal{A}$, defined as
\begin{equation}
S_\mathcal{A} = - \Tr_\mathcal{A} [ \rho_\mathcal{A} \ln (\rho_\mathcal{A}) ] \, .
\end{equation}
Indeed, using Eq.~\ref{eq:Method_densitymatrixcorrel}, we get
\begin{equation} \label{eq:Method_EEFromCorrelMatrix}
S_\mathcal{A} = - \Tr [ C_\mathcal{A} \ln C_\mathcal{A} + (1 - C_\mathcal{A}) \ln (1 - C_\mathcal{A}) ] \, .
\end{equation}
Fluctuations of the total charge $N_\mathcal{A}$ contained in the region $\mathcal{A}$, which are more easily accessible than $S_\mathcal{A}$ in experiments~\cite{Klich_2006,Klich_2009a},
can also serve to probe the system's nature. As for the entropy, the mean value, variance and all higher order cumulants of $N_\mathcal{A}$ can be obtained as a function of the correlation matrix eigenvalues.
To find compact expressions for those quantities, it is useful to introduce the generating function
\begin{equation}
f_\mathcal{A}(t) = \log \langle e^{tN_\mathcal{A}} \rangle = \Tr \log \left[ 1+(e^t-1) C_\mathcal{A} \right] \, .
\end{equation} 
For instance, the mean and variance of $N_\mathcal{A}$ are obtained as
\begin{align} \label{eq:Method_NAFromCorrelMatrix} 
\langle N_\mathcal{A} \rangle & = (\partial_t f_\mathcal{A})_{t=0} = \Tr (C_\mathcal{A}) \, , \\ 
V_\mathcal{A} & = \langle N_\mathcal{A}^2 \rangle - \langle N_\mathcal{A} \rangle^2 =  (\partial_t^2 f_\mathcal{A})_{t=0} = \Tr (C_\mathcal{A} - C_\mathcal{A}^2) \, . \notag
\end{align}
In the rest of the article, we rely on Eqs.~\ref{eq:Method_EEFromCorrelMatrix} and~\ref{eq:Method_NAFromCorrelMatrix} to compute the EE and charge fluctuations of lattice models hosting Dirac cones, either analytically or numerically.

\subsection{Illustrative example: the SSH model} \label{ssec:SSHmodel}

\begin{figure*}
\centering
\includegraphics[width=1.7\columnwidth]{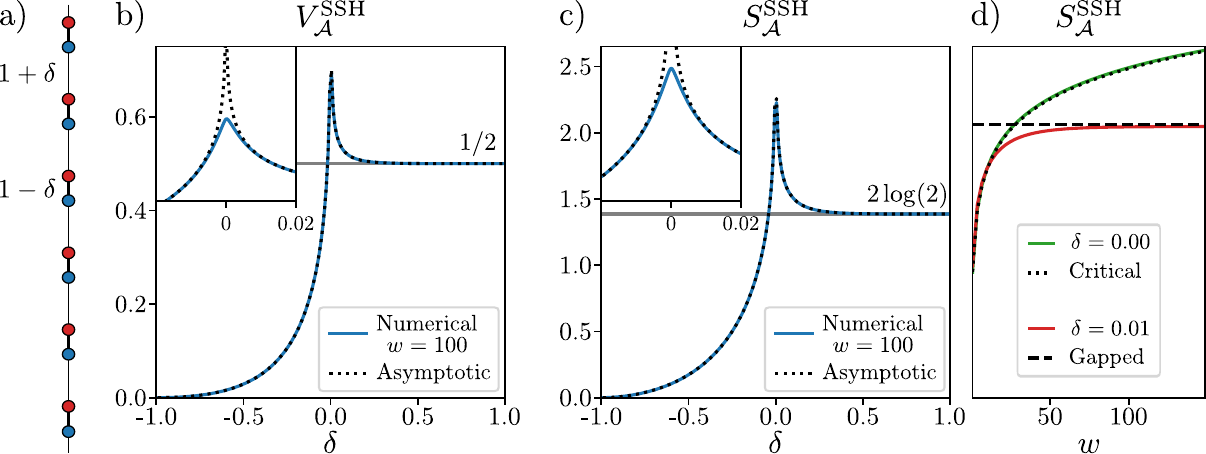}
\caption{
a) Schematic representation of the SSH chain with staggered hopping $1\pm\delta$.
b) Charge variance $V_{\mathcal{A}}^{\rm SSH}$ of the SSH model (blue)as a function of $\delta$ numerically evaluated for a segment of width $w=100$ in a finite but long chain ($N_x = 1024 \gg w$), compared to the asymptotic result Eq.~\ref{eq:SSH_Asymptotics_Variance} (dotted). The inset zooms in the region $|\delta| \leq 0.02$. 
c) Same as b for the EE $S_{\mathcal{A}}^{\rm SSH}$.
d) EE as a function of the number of unit cells $w$ in $\mathcal{A}$ at fixed $\delta$. For $\delta=0.01$ (red), \emph{i.e.}, slightly above the critical value, increasing the width $w$ above the correlation length yields converged results that match the asymptotic prediction. At the critical point $\delta = 0$ (or when $w<\xi_{\rm SSH}$ -- see text), the EE follows the Cardy-Calabrese relation $S_{\mathcal{A}}^{\rm SSH} = \log(w)/3$ (green).
}
\label{fig:SSHModelResults}
\end{figure*}

We illustrate the method outlined above on the SSH model~\cite{su1979solitons}, which describes spinless fermions with staggered hopping on a one-dimensional chain (see Fig.~\ref{fig:SSHModelResults}a). 
Its Fourier Hamiltonian is
\begin{equation} \begin{split}
& \tilde{h}_{\rm SSH}(q ) = \begin{bmatrix} 0 & f_{\rm SSH}(q, \delta) \\ f_{\rm SSH}^*(q, \delta) & 0 \end{bmatrix}  \\
& f_{\rm SSH}(q, \delta) = (1-\delta)  + (1+\delta) e^{iq} \, ,
\end{split} \end{equation}
with $- \pi < q \leq \pi$ a momentum label and $-1 \leq \delta \leq 1$ the dimensionless staggering amplitude. At half-filling, the lowest excitation above the ground state has energy $2|\delta|$. The corresponding correlation length is given by
\begin{equation}
\xi_{\rm SSH}(\delta) = \left|\log |\epsilon|\right|^{-1} \, , \quad \text{with: } \quad \epsilon = \frac{1-\delta}{1+\delta} \, .
\end{equation}
It diverges when $\delta \to 0$, where the model describes a half-filled and gapless system of spinless fermions with nearest-neighbor hopping.

Let us consider a region $\mathcal{A}(w)$ of $w$ consecutive unit cells, \textit{i.e.} of $2w$ consecutive sites. Its correlation matrix reads (see App.~\ref{app:CorrelationMatrix})
\begin{equation} \label{eq:SSH_CorelMatrix}
C_\mathcal{A}(r,r') = \int_{-\pi}^\pi \frac{{\rm d} q}{4\pi} e^{-iq(r-r')} \left[ 1- \frac{\tilde{h}_{\rm SSH}(q)}{|f_{\rm SSH}(q)|} \right] \, .
\end{equation}
The spectrum of the correlation matrix is known exactly in the limit of a very large interval $w \to \infty$. It has been obtained in Ref.~\cite{jin2004quantum,jin2007entropy} exploiting the fact that $C_\mathcal{A}$ is a block Toeplitz matrix, and using the Szeg\"o-Widom theorem. An alternative derivation based on the corner transfer matrix can be found in Ref.~\cite{Peschel_2004,Peschel_2009}. This spectrum is particularly simple :
\begin{align}
\lambda_m= \frac{1}{1+e^{m\pi\frac{I(k')}{I(k)}}}, \quad  \left\{ \begin{array}{ccc} m & \textrm{even if} & \delta > 0 \\  m &  \textrm{odd if} & \delta < 0 \end{array} \right..
\end{align}
with each eigenvalue $\lambda_m$ appearing twice,  $k = \min (\epsilon, 1/\epsilon)$, $k' = \sqrt{1-k^2}$ and $I(k) = \int_0^{\pi/2}  [1-k^2 \sin^2 \theta]^{-1/2} {\rm d}\theta$ the complete elliptic integral of the first kind. This remarkable formula leads to the following asymptotic limits of the EE~\cite{Peschel_2004,Eisler_2020}
\begin{widetext}
\begin{equation} \label{eq:SSH_Asymptotics}
S_{\mathcal{A}(w\to\infty)}^{\rm SSH} (\delta) = \begin{cases}
\frac{1}{3}\left[ \log\frac{4}{kk'} + (k^2-k'^2) \frac{2I(k)I(k')}{\pi} \right]  & {\rm if} \,\, \delta < 0\\ 
\frac{1}{3}\left[ \log\frac{k^2}{16k'} + (2-k^2) \frac{2I(k)I(k')}{\pi} \right] + 2\log 2 &  {\rm if} \,\, \delta > 0
\end{cases} \, , 
\end{equation}
\end{widetext}
Similarly, we can obtain the charge variance (see App.~\ref{app:SSHParticleVariance} for the detailed derivation) for $w \to \infty$
\begin{equation} \label{eq:SSH_Asymptotics_Variance}
V_{\mathcal{A}}^{\rm SSH} (\delta) = \begin{cases}
\frac{2I(k)E(k)}{\pi^2} - \frac{2k'^2 I(k)^2}{\pi^2}  & {\rm if} \,\, \delta < 0\\ 
\frac{2I(k)E(k)}{\pi^2} \,\, & {\rm if} \,\, \delta > 0
\end{cases} \, ,
\end{equation}
where $E(k)$ is the complete integral of the second kind $E(k) = \int_0^{\pi/2}  [1-k^2 \sin^2 \theta]^{1/2} {\rm d}\theta$.

These asymptotic limits are plotted in Fig.~\ref{fig:SSHModelResults}b-c as a dotted line. Their characteristic behaviour near the three particular points $\delta = -1,0,1$ can be intuitively understood. 
Let us first focus on $\delta = \pm 1$, for which one of the staggered tunneling coefficients is zero, and the system forms local independent dimers. The boundary $\partial \mathcal{A}$ either cuts two of these dimers into halves, leading to $S_{\mathcal{A}}^{\rm SSH}  = 2\log(2)$ and $V_{\mathcal{A}}^{\rm SSH} = 1/2$ for $\delta =1$, or does not divide any bound pairs, giving $S_{\mathcal{A}}^{\rm SSH}  = 0$ and $V_{\mathcal{A}}^{\rm SSH} = 0$ for $\delta =-1$. These are the two limits observed in Fig.~\ref{fig:SSHModelResults}b-c. 
Turning to $\delta$ close to zero, the system approaches its gapless point and the correlation length diverges as $\xi_{\rm SSH} \sim 1/2|\delta|$. 
When the latter is much larger than the lattice spacing, the universal properties of the model can be captured by a massive quantum field theory. For 1d systems, this yields the characteristic relation~\cite{calabrese2004entanglement,PhysRevB.82.012405} 
\begin{equation}
3 S_{\mathcal{A}}^{\rm SSH} \sim \pi^2  V_{\mathcal{A}}^{\rm SSH}  \sim \log (\xi_{\rm SSH}) \sim  - \log(2|\delta|),
\end{equation}
which correctly captures the logarithmic divergence of Eqs.~\ref{eq:SSH_Asymptotics} and~\ref{eq:SSH_Asymptotics_Variance} near $\delta=0$.

The explicit expression Eq.~\ref{eq:SSH_CorelMatrix} also allows direct access to the charge variance and the EE away from the asymptotic regime $w\to \infty$ by numerical diagonalization of $C_\mathcal{A}$. 
In Fig.~\ref{fig:SSHModelResults}b-c, this full-fledged numerical evaluation for a segment of length $w=100$ in a finite chain containing $N_x = 1024 \gg w$ unit cells is compared to the asymptotic result Eq.~\ref{eq:SSH_Asymptotics}. 
A perfect agreement is observed, except for $|\delta| < 0.01$ (inset), where we notice that the correlation length $\xi_{\rm SSH}(\delta) > w$ exceeds the size of $\mathcal{A}$.
In that region, the single-particle gap is smaller than the finite-size energy resolution $\sim 1/w$. 
Thus, the system restricted to $\mathcal{A}$ effectively behaves as a critical chain, and the EE should follow the Cardy-Calabrese relation with a central charge equal to one, \textit{i.e.} $S_{\mathcal{A}}^{\rm SSH} = \log(w)/3$~\cite{calabrese2004entanglement}. 
This is indeed what is observed at small $w$ in Fig.~\ref{fig:SSHModelResults}d.
If $w$ is increased above the $\xi_{\rm SSH}$, the thermodynamic limit is reached within region $\mathcal{A}$ and the asymptotic result Eq.~\ref{eq:SSH_Asymptotics} holds. 
This materializes in Fig.~\ref{fig:SSHModelResults}d as a departure from the Cardy-Calabrese formula and a saturation of the EE towards a constant. 
As explained above, the value of this constant approaches $\log (\xi_{\rm SSH})/3$ close to the critical point. 
In this saturated region, the EE does not depend on $w$ but rather scales with the size of the boundary $\partial \mathcal{A}$, which is a constant for a 1d chain, an example of the \textit{area law} that highlights the short-ranged correlations in gapped phases.

The SSH example provides an important insight, which will prove useful thereafter to understand our results: asymptotic results on the charge variance and the EE only apply when the typical size of $\mathcal{A}$ is greater than all other length-scales of the problem. In particular, the points where the system approaches criticality should be treated with great care. The SSH chain also offers the closed-form expression Eqs.~\ref{eq:SSH_Asymptotics} and~\ref{eq:SSH_Asymptotics_Variance} that we will use to quantitatively examine the properties of graphene in Sec.~\ref{sec:GrapheneExact}.

\section{Entanglement response to flux} \label{sec:GrapheneExact}

Entanglement properties are known to be a powerful probe to analyse the nature of quantum states. A promising idea to detect Dirac matter is to use the entanglement response to flux insertions~\cite{metlitski2009entanglement,chen2017two,Zhu2018EntanglementSO}. In this section, we review the field theory prediction to the scaling of the EE for a Dirac fermion, before presenting an exact calculation for the graphene lattice model.
\\

\subsection{Field theory prediction} \label{ssec:ContinuumPredictionFlux}

To put things in a broader context, let us first recall the behavior of EE for gapped phases. We focus on two-dimensional systems, and in this section we assume that the spatial region $\mathcal{A}$ has a smooth boundary, which we denote by $\partial \mathcal{A}$ (the boundary is sometimes referred to as the entangling surface).  The leading correction to the ubiquitous area law for a \emph{gapped} two-dimensional system is a universal constant correction $\gamma_{\textrm{topo}}$ dubbed topological entanglement entropy (TEE)~\cite{PhysRevLett.96.110404,Levin_2006}:
\begin{align}
\label{eq_generic_area_Law}
S_\mathcal{A} = \alpha L - \gamma_{\textrm{topo}} + O(L^{-1})  
\end{align}
In the above equation $L$ is the length of the boundary $\partial \mathcal{A}$. The TEE is universal in the renormalization group sense: it is insensitive to irrelevant perturbations, and thus only depends on the infrared, universal properties of the quantum phase under consideration. The infrared fixed-point of a gapped phase is described by a (possibly trivial) topological quantum field theory. The TEE $\gamma_{\textrm{topo}}$ only depends on the related topological data as well as the topology of the region $\mathcal{A}$, \emph{e.g.} the number of connected components of $\partial \mathcal{A}$: if the boundary $\partial \mathcal{A}$ has two components, then $\gamma$ is doubled. In particular the TEE vanishes for phases without intrinsic topological order.

\begin{figure}
\centering
\includegraphics[width=0.75\columnwidth]{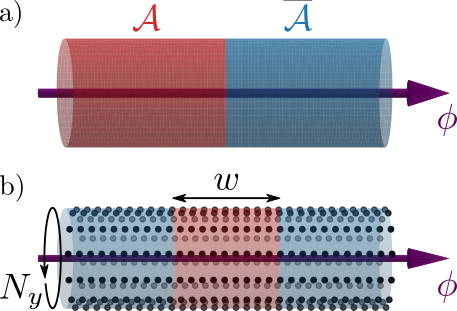}
\caption{a) For the two-cylinder EE, $\mathcal{A}$ is half of the an infinite cylinder threaded by a flux $\phi$. $\bar{\mathcal{A}}$ denotes the complement of $\mathcal{A}$. b) Our lattice calculations are done on a cylinder with $N_x \times N_y$ unit cells, assuming periodic boundary conditions along $y$, with $\mathcal{A}$ a slab of length $w$ unit cells in the $x$ direction.
}
\label{fig:CutDefinitions}
\end{figure}

In the case of critical Dirac matter, the infra-red theory capturing the low-energy universal properties is a 2+1 dimensional conformal field theory. Generically for gapless systems with an emerging conformal invariance, the area law is still expected to hold~\cite{Ryu_2006,Liu_2013,metlitski2009entanglement,chen2015scaling,PhysRevB.95.045148}:
\begin{align}
\label{eq_CFT_area_Law}
S_{\mathcal{A}} = \alpha L - \alpha_0 + O(L^{-1})  \,.  
\end{align}
where $\alpha_0$ is a constant (\emph{i.e.} scale-invariant) correction. Unlike the TEE $\gamma_{\textrm{topo}}$, which is insensitive to smooth deformations of the spatial region $\mathcal{A}$, the constant term $\alpha_0$ does depend on the shape of $\mathcal{A}$~\cite{Bueno_2017,chen2017two}. Furthermore, for theories with a U$(1)$ symmetry, $\alpha_0$ also depends on the magnetic flux~\cite{metlitski2009entanglement,chen2017two,Zhu2018EntanglementSO,Arias_2015,chen2017two,PhysRevB.95.045148}. Namely working on an infinite cylinder of perimeter $L$ and taking for region $\mathcal{A}$ a semi-infinite half-cylinder (see Fig.~\ref{fig:CutDefinitions}a), the EE for a single massless Dirac fermion reads~\cite{Arias_2015,chen2017two}
\begin{align}
\label{eq_sin_flux}
\alpha_0 =  \frac{1}{6} \log \left|2  \sin \frac{\phi}{2} \right|\,.  
\end{align}
where $\phi$ is the flux going through the cylinder, as depicted in Fig.~\ref{fig:CutDefinitions}a.The EE in this geometry has been coined two-cylinder entanglement entropy~\cite{chen2017two}. The presence of an exact zero mode at $\phi=0$ yields a divergence in Eq.~\ref{eq_sin_flux}. When the region $\mathcal{A}$ has a finite length $w$ along the cylinder direction, $\alpha_0$ is rather bounded by an amount proportional to $\log(w)$ when $\phi$ approaches zero, as hinted in Fig.~\ref{fig:SSHModelResults}d and highlighted in Ref.~\cite{chen2017two}.

It is rather tempting to exploit the non trivial dependence on shape and flux of $\alpha_0$ as a diagnostic tool to help identify the universality of a given critical model. But this raises the question of the robustness of this quantity. Being dimensionless, $\alpha_0$ does not depend on the short-distance cut-off. Based on this observation, it is generally assumed that $\alpha_0$ is a low energy property of the phase under consideration. In other words, $\alpha_0$ is typically believed to be universal in the renormalization group sense, that is insensitive to irrelevant perturbations. This question however is not fully resolved, and generally it is not known whether $\alpha_0$ can be reliably compared between field theories and lattice models. In the particular case of the flux response numerical evidence suggests that the behavior Eq.~\ref{eq_sin_flux} is indeed universal. In particular this signature has been used successfully in Refs.~\cite{Zhu2018EntanglementSO,Hu_2019} as a fingerprint for Dirac fermions in spin liquids and in the $\pi-$flux model. In order to further address this question, we consider the particular example of a graphene cylinder, hosting two Dirac cones.

\subsection{Exact lattice calculation for graphene} \label{ssec:GrapheneExactLatticeCalculation}

Focusing on graphene, we use dimensional reduction and the asymptotic results Eq.~\ref{eq:SSH_Asymptotics} to derive an exact formula for the corresponding EE of a segment of width $w \to \infty$. It exactly matches the continuum prediction Eq.~\ref{eq_sin_flux}, up to a factor $2$ accounting for the presence of two Dirac cones, and quantitatively agrees with numerical simulations. Moreover, our derivation can be easily generalized to any non-interacting tight-binding model hosting Dirac cones, thus providing a very strong argument in favor of the universality of the flux response Eq.~\ref{eq_sin_flux}. 

\subsubsection{Graphene as a collection of SSH chains} \label{sssec:DimReduc}

\begin{figure}
\centering
\includegraphics[width=\columnwidth]{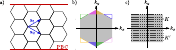}
\caption{
a) Honeycomb lattice with basis vectors $\vec{a}_1$ and $\vec{a}_2$ with periodic boundary condition along $y$. 
b) Corresponding first Brillouin zone, obtained as the Wigner–Seitz unit cell in momentum space. Translating the tips by a momentum lattice vector, as shown with colors gives the rectangular Brillouin zone used in the main text (c). In c), the allowed values of the momentum $q_y$ are shown with dashed lines, together with the two Dirac points $K$ and $K'$. 
}
\label{fig:BrillouinZone}
\end{figure}

Graphene can be modeled by a nearest-neighbor tight-binding model on the honeycomb lattice with Bloch Hamiltonian
\begin{equation}
\tilde{h}_{\rm G} (\vec{k}) = \begin{bmatrix} 0 & f^* \\ f & 0 \end{bmatrix} \, , \quad f = 1 + e^{i\vec{k}\cdot\vec{a}_1} + e^{i\vec{k}\cdot\vec{a}_2} \, ,
\end{equation}
with
\begin{equation}
\vec{a}_1 = \left( \frac{\sqrt{3}}{2} ,  \frac{1}{2} \right) \, , \quad \vec{a}_2 = \left( \frac{\sqrt{3}}{2} ,  - \frac{1}{2} \right) \, ,
\end{equation}
the two lattice basis vectors, and where the internal degree of freedom $\tau = A, B$ distinguishes the two inequivalent sites of the honeycomb unit cell (see Fig.~\ref{fig:BrillouinZone}a).
We assume that the system has $N_y$ unit cells and periodic boundary condition along the $y$ direction, \textit{i.e.} we identify any lattice site $\vec{r}$ with its translated $\vec{r}+N_y(\vec{a}_1-\vec{a}_2)$, which requires to consider a total perimeter of $N_y$.
Along the perpendicular direction $\vec{a}_1+\vec{a}_2$ pointing along the $x$-direction, we either consider an infinitely long cylinder for analytical purposes, or assumed periodic boundary conditions with a number of unit cells $N_x \gg N_y$ for numerical calculations.  
The momenta $\vec{k} = (k_x, k_y)$ satisfy
\begin{equation}
\vec{k} \cdot \vec{a}_j = \frac{\sqrt{3}k_x}{2} + (-1)^{1+j} \frac{k_y}{2} \, ,
\end{equation}
and can be restricted to a single Brillouin zone ($BZ$) that we choose rectangular and parametrized by the reduced momenta $q_x = \sqrt{3} k_x / 2  \in (-\pi, \pi]$ 
and $q_y \in (-\pi, \pi]$ (see Fig.~\ref{fig:BrillouinZone}b-c). 
The periodic boundary conditions along $y$ quantize the transverse momenta as
\begin{equation}
q_y = \frac{2p\pi +\phi}{N_y} \, , \quad p = -\left\lfloor{\frac{N_y-1}{2}}\right\rfloor , \cdots , \left\lfloor{\frac{N_y}{2}}\right\rfloor 
\end{equation}
where $\lfloor{x}\rfloor$ denotes the integer part of $x$. Here, $\phi$ denotes the flux threading the graphene along the cylinder axis, as sketched in Fig.~\ref{fig:CutDefinitions}b.
In terms of reduced momenta, the $K$ and $K'$ points, where the Dirac cones are located, lie at $(\pi, \pm K_y)$ with $K_y = 2\pi/3$, respectively.
They are only reached at zero flux if $N_y$ is divisible by 3, making the graphene cylinder (or nanotube) metallic.

\begin{figure}
\centering
\includegraphics[width=\columnwidth]{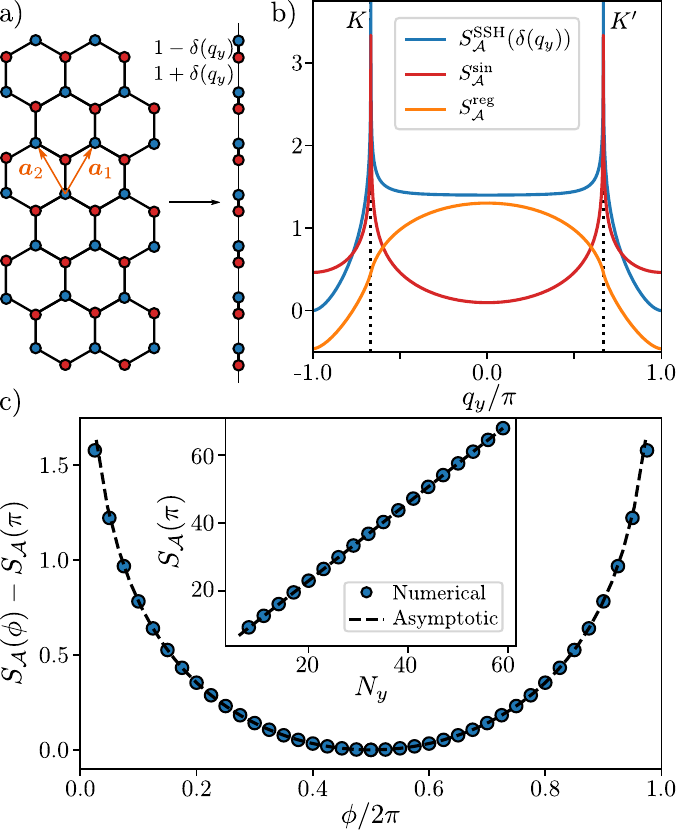}
\caption{
a) After Fourier transform along the $y$ direction, a graphene cylinder can be viewed as decoupled SSH chains with parameter $\delta(q_y)$ (see Eq.~\ref{eq:Graphene_StaggeredAmplitudeQy}). 
b) Using this dimensional reduction, the asymptotic EE of graphene can be evaluated exactly for a segment of length $w\to\infty$ preserving translational symmetry along $y$. We split the momentum-resolved EE into a singular and regular part. 
c) Comparison of the asymptotic results to the numerical evaluation of the graphene EE as a function of the flux $\phi$ for a segment of width $w=200$ on a cylinder of total size $N_x = 1024$ and $N_y=60$ with no fitting parameters. For convenience, we show the EE shifted by its value at $\phi=\pi$, \emph{i.e.}, $S_{\mathcal{A}}(\phi)-S_{\mathcal{A}}(\pi)$. In the inset, we probe the area law at $\phi=\pi$ by tuning $N_y$ up to $60$.}
\label{fig:ResultsGraphene}
\end{figure}

Having set up the necessary notations, we now recall that the graphene cylinders can be viewed as a collection of SSH chains, as schematically drawn in Fig.~\ref{fig:ResultsGraphene}a.
We first rewrite 
\begin{equation}
f = 1 + 2 \cos(q_y/2) e^{iq_x}  = Q(q_y) f_{\rm SSH} [q_x, \delta (q_y)] \, ,
\end{equation}
with $Q(q_y) = 2 [2\cos(q_y/2) + 1]^{-1}$ and 
\begin{equation} \label{eq:Graphene_StaggeredAmplitudeQy}
\delta (q_y) =  \frac{2\cos(q_y/2)-1}{2\cos(q_y/2)+1} \, .
\end{equation}
Hence, we assign for each value of $q_y$ an effective SSH chain in the $x$ direction with a staggering parameter $\delta (q_y)$.

This representation as a collection of independent SSH chains allows to evaluate EEs and charge fluctuations for graphene tubes.
Consider for region $\mathcal{A}$ a slab of the cylinder of length $w$ along $x$ (see Fig.~\ref{fig:CutDefinitions}b). 
At zero energy, all states with negative energies are filled, and the positive prefactor $Q(q_y)$ could be replaced by one when computing the correlation matrix (see Eq.~\ref{eq:Method_DefinitionCorrelMatrix}). 
The EE of the graphene cylinder thus reads
\begin{equation} \label{eq:Graphene_EESumSSH}
S_\mathcal{A} = \sum_{q_y} S_\mathcal{A}^{\rm SSH} [\delta(q_y)] \, .
\end{equation}

\subsubsection{Asymptotic flux dependence}\label{sec:EEgrapheneasymptotic}

Using the asymptotic result for the SSH chain (Eq.~\ref{eq:SSH_Asymptotics}), we now infer the expression of the graphene EE in the $w \to \infty$ limit.
Because the SSH EE diverges at the Dirac closing points, where $\delta (\pm K_y) = 0$, we decompose it into a regular and singular part 
\begin{equation}
S_\mathcal{A}^{\rm SSH} [\delta(q_y)] = S_\mathcal{A}^{\rm sin} (q_y) + S_\mathcal{A}^{\rm reg} (q_y) \, .
\end{equation}
From the known behaviour of the elliptic integral $I(k)$~\cite{DLMF} and the definition of $\delta(k_y)$ given by Eq.~\ref{eq:Graphene_StaggeredAmplitudeQy}, we obtain the singular part
\begin{equation} \label{eq:Graphene_singularPart}
S_\mathcal{A}^{\rm sin} (q_y) = - \frac{1}{3} \log \left| \sin\left( \frac{q_y -K_y}{2} \right) \sin\left( \frac{q_y+K_y}{2} \right) \right| .
\end{equation}
The regular part $S_\mathcal{A}^{\rm reg} = S_\mathcal{A}^{\rm SSH} - S_\mathcal{A}^{\rm sin} $ follows from Eq.~\ref{eq:SSH_Asymptotics}. 
These two contributions are shown in Fig.~\ref{fig:ResultsGraphene}b, where the divergence of the momentum-resolved EE clearly appears near the $K$ and $K'$ points.

Using twice the identity
\begin{equation}
\prod_{p=0}^{N_y-1} \sin \left(\frac{p\pi}{N_y} + x\right) = \frac{\sin (N_y x)}{2^{N_y-1}} 
\end{equation}
for $x = \frac{\phi - 2 \pi \lfloor (N_y-1)/2 \rfloor}{2N_y} \pm \frac{K_y}{2}$, we find the following contribution of $S_{\mathcal{A}}^{\rm sin}$ to the graphene EE:
\begin{align}
\sum_{q_y} S_{\mathcal{A}}^{\rm sin} (q_y) = & \frac{2 N_y}{3} \log(2) \\
& - \frac{1}{3} \sum_{K\in\{\pm K_y\}} \log \left| 2 \sin \left[ \frac{\phi - N_y K}{2} \right] \right| . \notag
\end{align}
The first term of the right hand side contributes as an area law term, while the second one is exactly the expected flux dependence for two Dirac cones located at $\pm K_y$ (see Eq.~\ref{eq_sin_flux}).
Because $S_{\mathcal{A}}^{\rm reg}$ is periodic in $q_y$ and sufficiently smooth, we can replace the sum by an integral up to exponentially small correction in $N_y$ through~\cite{TrapezoidalRulePeriodicFunction}
\begin{equation} \label{eq:EEGrapheneLargeW_bis}
\sum_{q_y} S_\mathcal{A}^{\rm reg} (q_y) = \frac{N_y}{2\pi} \int_{-\pi}^{\pi} S_\mathcal{A}^{\rm reg} (q) {\rm d}q + \mathcal{O} \left( e^{-\kappa N_y} \right) \, ,
\end{equation}
with $\kappa >0$.

\subsubsection{Summary and numerical checks}\label{sec:EEgrapheneflux}

Altogether, we find that 
\begin{equation}
S_{\mathcal{A}(w\to \infty)} = \alpha N_y - \frac{1}{3}  \log \left| \prod_{K = \pm K_y} \!\!\! 2 \sin \left[ \frac{\phi-N_y K}{2} \right] \right| ,\label{eq:EEGrapheneLargeW}
\end{equation}
with $\alpha = \frac{2 \log 2}{3} + \int_{-\pi}^{\pi} S_\mathcal{A}^{\rm reg} (q) \frac{{\rm d}q}{2\pi}$. This proves that, for $w$ sufficiently large, the EE of a graphene tube quantitatively matches the continuum prediction Eq.~\ref{eq_sin_flux}. 
We compare this asymptotic prediction to the numerical results obtained for a region $\mathcal{A}$ of width $w = 200$ in Fig.~\ref{fig:ResultsGraphene}, where we stress that no fitting parameters are used since the integral in $\alpha$ is evaluated numerically.
A perfect agreement is found between the numerical and the asymptotic results, except near the gap closing points $\phi = 0, 2\pi$. 
This is expected since the correlation length of the SSH chain with $\delta(\pm K_y)=0$ diverges, which forbids the use of the asymptotic results for a finite $w$ as considered in a numerical calculation (see discussion in Sec.~\ref{ssec:SSHmodel}). 

\subsection{Charge fluctuations} \label{ssec:GrapheneChargeFluctuations}

Although our discussion has been so far focused on the EE, it naturally extends to the charge fluctuations.
Indeed, the reduction of the graphene cylinder to a collection of SSH chains allows us to express the variance $V_{\mathcal{A}}$ as a sum of variances of the form Eq.~\ref{eq:SSH_Asymptotics_Variance}. 
More precisely, we introduce $n_{\mathcal{A}}(q_y)$ the number of particles localized in region $\mathcal{A}$ with transverse momentum $q_y$, and find that
\begin{equation}
f_\mathcal{A}(t) = \log \langle e^{t \sum_{q_y} n_{\mathcal{A}} (q_y)} \rangle = \sum_{q_y} \log \langle e^{t n_{\mathcal{A}} (q_y)} \rangle \, .
\end{equation}
All cumulants of $N_{\mathcal{A}}$ inherit the additivity of the generating function $f_{\mathcal{A}}$, and the charge variance of the graphene slab becomes
\begin{equation}
V_{\mathcal{A}} = \sum_{q_y} V_{\mathcal{A}}^{\rm SSH} [\delta(q_y)] .
\end{equation}

Following Sec.~\ref{sec:EEgrapheneasymptotic}, we then split the asymptotic expression of the variance given in Eq.~\ref{eq:SSH_Asymptotics_Variance} into a singular and a regular part  $V_{\mathcal{A}}^{\rm SSH} = V_{\mathcal{A}}^{\rm reg} + V_{\mathcal{A}}^{\rm sin}$, with $V_{\mathcal{A}}^{\rm sin} = (\pi^2 / 3) S_{\mathcal{A}}^{\rm sin}$ as shown in App.~\ref{app:SSHParticleVariance}. Summations over $q_y$ are performed identically to the EE.
We thus obtain for the charge variance 
\begin{equation} \label{eq:FluxDependenceOfVariance}
V_{\mathcal{A}} = \beta N_y - \frac{1}{\pi^2} \log \left| \prod_{K=\pm K_y} 2\sin\left[ \frac{\phi-N_yK}{2} \right] \right| ,
\end{equation}
with $\beta = 2\log 2 /\pi^2 + \int_{-\pi}^\pi V_{\mathcal{A}}^{\rm reg} \frac{{\rm d}q_y}{2\pi}$ analogous to $\alpha$ in Eq.~\ref{eq:EEGrapheneLargeW}.
We will show in Sec.~\ref{sec:NumericalChargeFluctuations}, how this expression accurately captures the direct numerical evaluation of the charge variance.

We can repeat a similar argument to express higher order cumulants 
\begin{equation}
\kappa_n = \left. \partial_t^n   f_\mathcal{A}(t) \right|_{t=0} , \quad n > 2 ,
\end{equation}
as a function of their counterparts in the SSH model
\begin{equation}
\kappa_n = \sum_{q_y} \kappa_n^{\rm SSH} [\delta (q_y)] .
\end{equation}
As shown in App.~\ref{app:SSHParticleVariance}, the latter are regular at $\delta=0$: charge fluctuations of 1d critical systems are Gaussian~\cite{Abanov_2011}. More precisely, only the variance increases proportionally to $\log w$ when the length of the interval $w$ increases, while all higher order cumulants eventually saturates to constant values. 
Because the singular part vanishes, we find that 
\begin{equation}
\kappa_n = \frac{N_y}{2\pi} \int_{-\pi}^\pi \kappa_n^{\rm SSH} [\delta (q)] \; {\rm d}q ,
\end{equation}
up to exponentially small corrections, as in Eq.~\ref{eq:EEGrapheneLargeW_bis} for the regular part of the entropy.
Therefore, higher order cumulants only exhibit exponentially small flux dependent corrections to the area law. For this reason, we will only consider the variance in the rest of this article.

\section{Extension and numerical results}\label{sec_extension_num}

In this section, we show that the exact results derived for graphene can be easily generalized to any non-interacting tight-binding model hosting Dirac cones, thus providing a very strong argument in favor of the universality of the flux response Eq.~\ref{eq_sin_flux}. Furthermore, we highlight that the typical flux-dependence of Dirac cones does not come from the particular choice of the region $\mathcal{A}$ used in our derivation. Indeed, it is observed as long as the region $\mathcal{A}$ winds around the cylinder. We provide similar evidence for the charge variance.

\subsection{Other models} \label{ssec:OtherModels}

The separation of the momentum-resolved entropy into a singular and regular part offers simple generalizations to other models and lattices.
Indeed, the regular part only contributes to the non-universal area law coefficient $\alpha$, while all the flux dependence stems from the singular part.
The latter is free from any microscopic details. Indeed, it models the logarithmic divergence of the EE $\log (\xi)/3$ near each of the Dirac points, where the correlation length is given by $\xi(q_y) \sim |q_y-K|^{-1}$ due to the characteristic linear dispersion relation of the Dirac cone.
The formula Eq.~\ref{eq:Graphene_singularPart} can be straightforwardly extended to any model with $N_D$ Dirac cones located at $K_{y,1}, \cdots , K_{y,N_D}$ along $q_y$, and yields the following flux dependence for the EE
\begin{equation}
\label{eq_generic_flux}
S_{\mathcal{A}} = \alpha N_y - \frac{1}{3} \log \left| \prod_{i=1}^{N_D} 2 \sin\left[ \frac{\phi-\phi_i}{2} \right] \right| \, .
\end{equation}
Here, $\phi_i = N_y K_{y,i}$ is the flux at which  one of the finite size momenta $q_y=(2\pi p +\phi)/N_y$ reaches $K_{y,i}$. This flux dependence of the EE appeared as an ansatz in Ref.~\cite{Zhu2018EntanglementSO}. The above argument ascertains that this formula is indeed valid for any (non-interacting) lattice model hosting Dirac cones.

\begin{figure}
\centering
\includegraphics[width=\columnwidth]{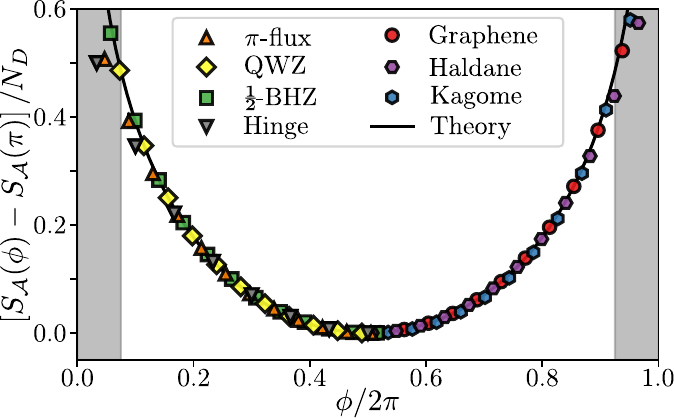}
\caption{
Flux dependence of the EE (shifted by its value at $\phi=\pi$, and per Dirac cone) for six distinct 2d models hosting $N_D=1$ or $2$ Dirac modes (see App.~\ref{app:TwoDimFerionicModels}), with $\mathcal{A}$ a slab of length $w=100$ on a cylinder with total dimensions $N_x=512$ and $N_y=60$.
They all follow the expected flux dependence given by Eq.~\ref{eq_sin_flux} (black lines).
We also observe the same behaviour for the surface mode of a 3d model labeled "hinge", which holds one surface Dirac mode on its top and bottom surface (see text and App.~\ref{app:HingeModel} for more details). For convenience, models with a square Bravais lattice are shown for $\phi \leq \pi$, whereas those defined on the honeycomb lattice are depicted for $\phi > \pi$ (the results being symmetric around $\phi=\pi$).
The grey areas close to $\phi=0,2\pi$ correspond to the cases where one of the momenta is getting close to (at least) one Dirac singularity. There, the finite value of $w$ leads to deviation to the asymptotic expression of Eq.~\ref{eq_sin_flux}. 
}
\label{fig:AllModelCollapse}
\end{figure}

We check this statement numerically by considering six different 2d models hosting either $N_D=1$ or $2$ Dirac cones (see App.\ref{app:TwoDimFerionicModels}) and the surface Dirac mode of a 3d model (see App.~\ref{app:HingeModel}). As suggested by our derivation in Sec.~\ref{ssec:GrapheneExactLatticeCalculation}, the results presented in Fig.~\ref{fig:AllModelCollapse} hint that the lattice regularization has little effect on the flux dependence of the EE, which always follows the prediction Eq.~\ref{eq_sin_flux}.
We also observe the same behaviour in a 3d model labeled "hinge" in Fig.~\ref{fig:AllModelCollapse}, which holds one surface Dirac mode on its top and bottom surface (see App.~\ref{app:HingeModel}). 
In Fig.~\ref{fig:AllModelCollapse}, we choose 3d bulk of dimension $(N_x ,N_y , N_z) = (100,20,60)$ and a region $\mathcal{A}$ of size $(30,20,20)$ starting from the top surface, in order to only enclose the Dirac mode from the upper surface.

As in the graphene case, we observe substantial corrections to Eq.~\ref{eq_sin_flux} when $\phi$ is tuned such that one of the momenta $q_y$ hits (or getting close to) the center of a Dirac cone, which occurs for $\phi = 0,2\pi$ in Fig.~\ref{fig:AllModelCollapse}.
In the illustrative example of Sec.~\ref{ssec:SSHmodel}, we observed that asymptotic results for the EE only hold when $w$ is greater than the largest correlation length of the system.
This largest correlation length is of order $\sim N_y / \phi$ originating from the finite size gap close to the Dirac cone band closing.
Hence, finite-size numerical simulation necessarily fail to capture the thermodynamic behavior Eq.~\ref{eq_sin_flux} when $\phi$ is too close to 0 or $2\pi$, where we instead anticipate non-universal lattice-dominated physics.

\subsection{Topology of the sub-region $\mathcal{A}$} \label{ssec:TopologyRegionA}

While we have heavily relied on the translational symmetry of the region $\mathcal{A}$ along the cylinder perimeter to verify the flux-dependence Eq.~\ref{eq_sin_flux} in lattice models, the slab geometry is not the only one where the characteristic flux response of Dirac cones appears. 
We now present numerical evidence showing that the same behaviour arises if and only if the region $\mathcal{A}$ wraps around the cylinder. 
We perform all our simulations on the $\sfrac{1}{2}$-BHZ model~\cite{bernevig2006quantum}, which describes tunneling of spin-polarized fermions on a square lattice with two orbitals per unit cells. 
The tunneling phases between the orbitals and the on-site potential difference are tuned such that the system hosts a single Dirac cone at the center of the Brillouin zone (see App.~\ref{app:TwoDimFerionicModels}).

\begin{figure}
\centering
\includegraphics[width=\columnwidth]{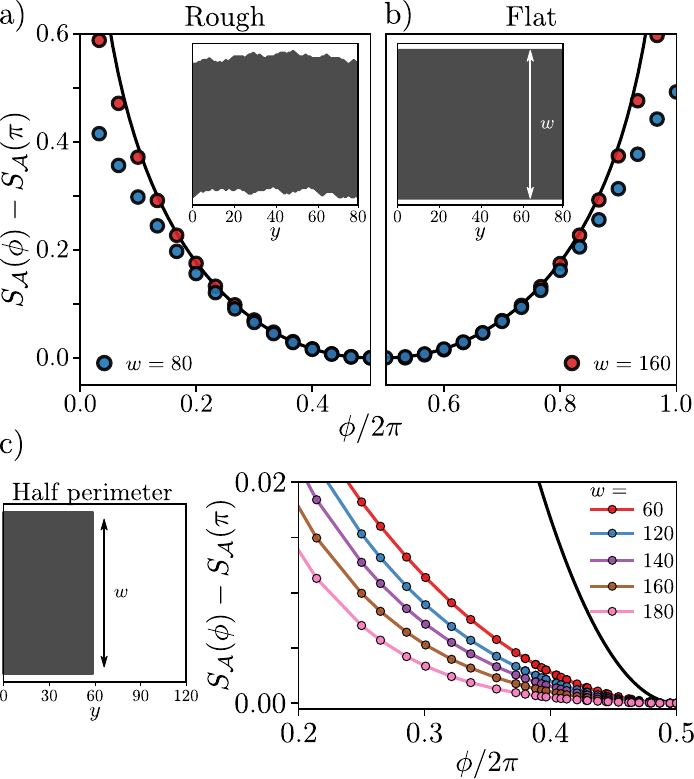}
\caption{
Flux dependence of the EE in the \sfrac{1}{2}-BHZ model for different geometry of $\mathcal{A}$. For convenience, we show the EE shifted by its value at $\phi=\pi$, \emph{i.e.}, $S_{\mathcal{A}}(\phi)-S_{\mathcal{A}}(\pi)$. In a) and b), the entangling region fully wraps around the cylinder with perimeter $N_y=80$ and total length $N_x = 1024$. It either has rough (a) or flat (b) boundaries.  In c), the region $\mathcal{A}$ only cover half of the cylinder perimeter $N_y=120$, as shown in the left panel. The right panel of c) gives the flux dependence of the EE 
With a larger $w$ that mitigates finite-size effects, a) and b) converge toward the continuum prediction Eq.~\ref{eq_sin_flux} (black line), while the c) decreases to zero.
This suggests that the typical flux dependence of Dirac cones is observed if and only if the region $\mathcal{A}$ wraps around the cylinder, \textit{i.e.} it depends on its topology.
}
\label{fig:RoughCylinder}
\end{figure}

We first consider a region $\mathcal{A}$ winding around a cylinder of perimeter $N_y=80$, with boundary surfaces that break translational symmetry along the $y$ direction, as shown in the inset of Fig.~\ref{fig:RoughCylinder}a. 
We numerically generated $\mathcal{A}$ with two random walks along the cylinder perimeter and returning to the origin that we separated by a mean distance $w$. 
The EE extracted as a function of the flux $\phi$ is shown in Fig.~\ref{fig:RoughCylinder}a.
It follows the continuum expectation Eq.~\ref{eq_sin_flux}, up to small corrections that we attribute to finite size effects. 
Indeed, they decrease with larger $w$, in agreement with the discussion of Sec.~\ref{sec:GrapheneExact}. 
Moreover, these discrepancies are similar in magnitude for the rough surface $\mathcal{A}$ and for a slab with flat edges shown in Fig.~\ref{fig:RoughCylinder}b for comparison. 
This indicates that, up to finite-size corrections, the typical flux-dependence of Dirac cones appears when the region $\mathcal{A}$ wraps around the cylinder, irrespective of the boundary translational symmetry or its smoothness.

On the contrary, the flux response is not observed if $\mathcal{A}$ does not wind around the cylinder, irrespective of the shape of the boundary. 
This can be seen in Fig.~\ref{fig:RoughCylinder}c, where the flux-dependence of the EE is presented for a rectangular patch of size $w \times (N_y/2)$, which only covers half of the cylinder perimeter.
The variation upon inserting the flux $\phi$ is drastically reduced in this geometry compared to the previous case, by a factor of about 10 (see Fig.~\ref{fig:RoughCylinder}).
While we see that the EE further reduces with larger $w$, we cannot reliably affirm that it converges to zero from our numerical data, especially when the flux $\phi$ is close to 0 or $2\pi$, where one of the finite size momenta reaches the Dirac cone. 
Nevertheless, our numerical results show a clear departure from Eq.~\ref{eq_sin_flux} when the region $\mathcal{A}$ does not wrap around the cylinder. This numerical evidence points out that the flux response of the Dirac cones only emerges when the entangling region winds around the cylinder.

\subsection{Charge fluctuations}\label{sec:NumericalChargeFluctuations}

The asymptotic expression for the charge fluctuations $V_{\mathcal{A}}$ of a long ($w\to\infty$) slab of graphene was derived in Sec.~\ref{ssec:GrapheneChargeFluctuations}. Similar to the EE in Secs.~\ref{ssec:OtherModels} and~\ref{ssec:TopologyRegionA}, we consider the generalization to other models hosting Dirac cones or changes in the topology of the entangling region. The arguments put forward for the EE also apply to the charge variance. We thus expect a similar universality of the flux dependence to hold true in that context.

\begin{figure}
\centering
\includegraphics[width=\columnwidth]{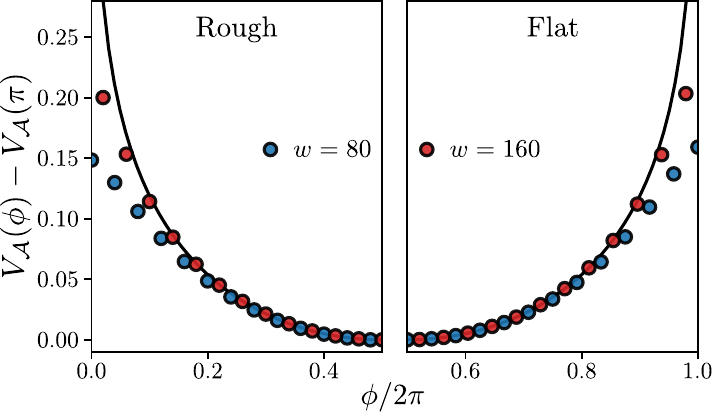}
\caption{Charge variance measured with respect to its value at $\phi=\pi$ for the \sfrac{1}{2}-BHZ model tuned with a single Dirac cone. We use the same geometries as Fig.~\ref{fig:RoughCylinder}a (for the rough entanglement surface) and Fig.~\ref{fig:RoughCylinder}b (for the flat entanglement surface), including the two values of $w$, namely $w=80$ (in blue) and $w=160$ (in red).
The solid black line is the asymptotic prediction of Eq.~\ref{eq:FluxDependenceOfVariance}.}
\label{fig:ParticleVrianceNumerics}
\end{figure}

We numerically test the predicted flux dependence of the charge variance Eq.~\ref{eq:FluxDependenceOfVariance} derived in Sec.~\ref{ssec:GrapheneChargeFluctuations}. For pedagogical purposes, we solely focus on the \sfrac{1}{2}-BHZ model. The conclusions hold true for all the models, including the surface of the 3d model considered in Section.~\ref{ssec:OtherModels} (see App.~\ref{app:HingeModel}). We consider for the entangling region $\mathcal{A}$ a slab of cylinder with either flat or rough edges, as described in Sec.~\ref{ssec:TopologyRegionA}. 
Our numerical results shown in Fig.~\ref{fig:ParticleVrianceNumerics} very well agree with the asymptotic predictions given by Eq.\ref{eq:FluxDependenceOfVariance} for $\phi$ not too close to 0 or $2\pi$, as expected from previous discussion on finite size effects.

\section{Kitaev-Preskill Subtraction Scheme}\label{sec_KP}

Up to here, we have mainly focused on spatial regions $\mathcal{A}$ with smooth boundaries. Avoiding sharp angles in $\partial \mathcal{A}$ has required us to only consider entangling regions that wind around the entire system. The area of such regions unfortunately grows extensively with one of the total system's dimension, making it hard to obtain reliable numerical results for analytically intractable models. In most case, one must therefore deal with the presence of sharp angles on the boundary $\partial \mathcal{A}$ in order to perform calculations on the lattice. 

While corners add extra terms to the EE even for gapped phases, subtraction schemes have been designed to eliminate their effects for gapped phases of matter and to yield universal results characterizing the system. In this section, we briefly review the corner contributions to the EE, and the most-used subtraction scheme. Then, we show that, in the presence of Dirac cones, subtractions schemes provide non-universal results that depend both on the lattice model and the specific cuts chosen to perform the subtractions.

\subsection{Corner contributions}\label{sec:CornerContributions}

Corrections to the leading behavior of the EE are sensitive to the geometry of the region $\mathcal{A}$, and in particular to the presence of corners in the boundary $\partial \mathcal{A}$. Before moving on to the case of quantum critical points, let us first recall how corners affect the EE of gapped phases, and how this can be remedied via a subtraction scheme.\\

In the absence of corners (\emph{i.e.} for a smooth entangling surface $\partial \mathcal{A}$), the correction to the ubiquitous area law for a \emph{gapped} two-dimensional system is the universal topological EE. 
For a non-smooth entangling surface $\partial \mathcal{A}$, additional non-universal constant terms coming from each corner spoil the above behavior\cite{PhysRevLett.96.110404,Rodriguez_2010}:
\begin{align}
S_{\mathcal{A}} = \alpha L - \gamma - \sum_{\textrm{corners i}} \gamma(\theta_i) + O(L^{-1})  \,.  
\end{align}
These corner contributions are encoded in a function $\gamma(\theta)$ of the the corner opening angle $\theta$. Naively one might expect the corner contribution to be universal, as it does not depend on the short-distance cut-off. However this is not so clear, since angle contributions are of ultraviolet origin. An argument against the universality of the corner terms is that they are not captured by the infrared topological quantum field theory, since the lack of a metric rules out the possibility to have any angle-dependent quantity. To put things short, angles are not topological invariant. In the case of the charge variance, the same scaling holds and the corner contributions are known explicitly~\cite{estienne2021cornering}.

These corner terms are potentially an issue for numerical calculations: on the lattice sharp corners are commonplace, making the direct extraction of the TEE $\gamma$ from a single EE computation hazardous. A workaround is to use a subtraction scheme~\cite{PhysRevLett.96.110404,Levin_2006}  - namely a certain linear combination of entanglement entropies for some well chosen regions sharing part of their boundaries - in which both the linear area law term and the corner contributions cancel out, leaving out the TEE $\gamma$. Key to this cancellation is the following relation obeyed by the corner functions
\begin{align}
\gamma(\theta) = \gamma(2\pi - \theta) 
\end{align}
which simply stems from the fact that $S_\mathcal{A} = S_{\bar{\mathcal{A}}}$. \\

For quantum critical points such as graphene at half-filling, the corner corrections exhibit a different scaling.  These have been discussed in the context of (2+1)-dimensional conformal field theories in Refs.~\cite{Fradkin_2006,Casini_2007,Hirata_2007,CASINI2009594,PhysRevLett.110.135702,Kallin_2014,PhysRevB.90.235106,PhysRevLett.115.021602,Bueno_2015,Bueno_2015b,PhysRevB.93.045131,Faulkner_2016,PhysRevB.95.045148,Bueno_2019} and in particular for Dirac fermions in Refs.~\cite{CASINI2009594,Casini_2009,Helmes_2016}. Trihedral corners for three-dimensional Dirac fermsions have also been consider in \cite{Bednik_2019}. For a conformal field theory, the EE behaves as 
\begin{align} \label{eq_generic_corner_diraccones}
S_{\mathcal{A}} = \alpha L - \sum_{\textrm{corners i}} a(\theta_i) \log L + O(L^{0})  \,.  
\end{align}
As opposed to the gapped case, the critical corner function $a(\theta)$ is universal. This is rather reasonable given that angles are conformal invariants. \\

As for the gapped case, these corner contributions spoil the constant term. Indeed upon changing the short-distance cutoff, or equivalently changing the unit in which lengths are measured, the logarithmic terms yield additional constant terms. But the situation here is even more muddled: subtraction schemes fail to eliminate corner contributions and it is no longer possible to extract the universal contribution $\alpha_0$ of Eq.~\ref{eq_CFT_area_Law}.

\subsection{Numerical results}

\begin{figure}
\centering
\includegraphics[width=\columnwidth]{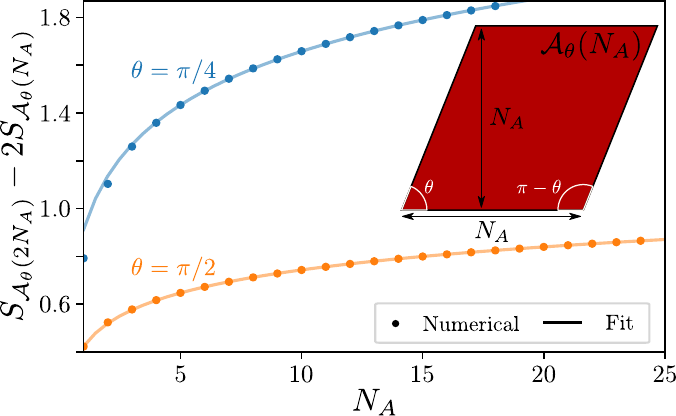}
\caption{Subtraction scheme to isolate corner contributions of Dirac cones in the \sfrac{1}{2}-BHZ model (see App.~\ref{app:TwoDimFerionicModels}) for $\theta = \pi/4$ and $\pi/2$, on a $1024\times1024$ finite-size lattice. The numerical data perfectly agree with the expectation of a dominant logarithmic scaling (solid lines show fits to Eq.~\ref{eq_subtractionscheme_logcorners} with $u$ and $v$ as fitting parameters). 
}
\label{fig:LogarithmCornersForDirac}
\end{figure}

We first present evidence of the logarithmic corrections to the EE due to corners in $\partial \mathcal{A}$. Let us denote as $\mathcal{A}_\theta(N_A)$ a parallelogram with base and height $N_A$, and angles $\theta$ and  $\pi - \theta$, as shown in Fig.~\ref{fig:LogarithmCornersForDirac}. 
According to Eq.~\ref{eq_generic_corner_diraccones}, corner contributions in the presence of a Dirac cone can be extracted through
\begin{equation} \label{eq_subtractionscheme_logcorners}
S_{\mathcal{A}_\theta(2N_A)} - 2 S_{\mathcal{A}_\theta(N_A)} = u \log N_A + v \, ,
\end{equation}
with $u=2[ a(\theta) + a(\pi-\theta)]$ and $v$ a non-universal constant. 
Numerical extractions of $S_{\mathcal{A}_\theta(2N_A)} - 2 S_{\mathcal{A}_\theta(N_A)}$ for the \sfrac{1}{2}-BHZ model hosting one Dirac cone (see App.~\ref{app:TwoDimFerionicModels}) are very well captured by this logarithmic behavior, as shown in Fig.~\ref{fig:LogarithmCornersForDirac}. 
We have observed the same logarithmic scaling in all the models and shapes considered.

Furthermore, least-square fitting allows to extract values such as $[a(\pi/4) + a(3\pi/4)] \simeq 0.0831$, which agree with the expectation 0.0826 for continuum theories~\cite{Helmes_2016}. 
This numerical check confirms the corner contribution of Dirac cones to the EE, which has already been observed in Refs.~\cite{Helmes_2016,herviou2019bipartite}.

\begin{figure}
\centering
\includegraphics[width=\columnwidth]{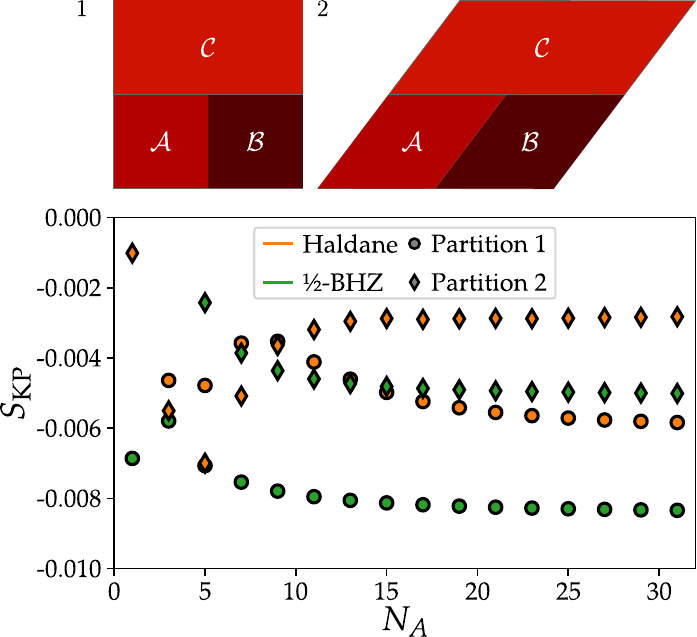}
\caption{$S_{\rm KP}$ for two lattice models hosting a single Dirac, respectively on the honeycomb (Haldane) and square lattice ($\sfrac{1}{2}$-BHZ) -- see App.~\ref{app:TwoDimFerionicModels} -- with total dimensions $N_x = N_y = 512$. Two different choices of regions $\mathcal{A}-\mathcal{B}-\mathcal{C}$ are considered (top), see text for more details. 
$S_{\rm KP}$ converges to a constant as $N_A$ increases, but this constant depends both on lattice details and the specific partition chosen. $S_{\rm KP}$ does not provide a universal characterization of models with Dirac cones. 
}
\label{fig:NonUniversalEE}
\end{figure}

The universality of the critical corner function $a(\theta)$ suggests that we could hope for its extraction as a numerical signature of the Dirac physics, using for instance a Kitaev-Preskill subtraction scheme. Unfortunately, the logarithmic factor of $a(\theta)$ in Eq.~\ref{eq_generic_corner_diraccones} spoils such a scheme with non-universal contributions. We illustrate this fact with the numerical evaluation of 
\begin{equation}
S_{\rm KP} = S_{\mathcal{A} \mathcal{B}} + S_{\mathcal{B} \mathcal{C}} + S_{\mathcal{C} \mathcal{A}} - S_{\mathcal{A}} - S_{\mathcal{B}} - S_{\mathcal{C}} - S_{\mathcal{A}\mathcal{B}\mathcal{C}} \, ,
\end{equation}
for two lattice models hosting a single Dirac, respectively on the honeycomb and square lattice -- see App.~\ref{app:TwoDimFerionicModels}. 
The regions $\mathcal{A}$, $\mathcal{B}$ and $\mathcal{C}$ are defined in Figs.~\ref{fig:NonUniversalEE}a-b. 
In both cases, $\mathcal{A}$ and $\mathcal{B}$ are shifted and adjacent copies of $\mathcal{A}_\theta(N_A)$ and $\mathcal{C} = \mathcal{A}_\theta(2N_A) \backslash (\mathcal{A}\cup \mathcal{B})$. 
Figs.~\ref{fig:NonUniversalEE}a and~b only differ by the value of $\theta$, either equal to $\pi/2$ for the first partition or to $\pi/4$ for the second one. 
The numerical results of Fig.~\ref{fig:NonUniversalEE}c show that $S_{\rm KP}$ converges to a constant as $N_A$ increases. 
However, this constant is not the same for both models, nor for the different choice of regions $\mathcal{A}-\mathcal{B}-\mathcal{C}$ for the same model. 
Hence, this lattice, model and geometry-dependent constant cannot be used as a universal probe of the presence of Dirac cone. 
The universality breakdown of $S_{\rm KP}$ comes from the constant corrections to the area law and logarithmic corner contribution in Eq.~\ref{eq_generic_corner_diraccones}, which are both model and geometry dependent. 
The ratio and other simple functions of the asymptotic values for a given model also appear to be non-universal.

\section{Conclusion}

In this paper we discussed the universal signature of Dirac physics in the EE and the charge fluctuations. For that purpose, we studied several tight-binding models whose low-energy physics is captured by Dirac fermions. In addition to the numerical investigation, we provided an analytical derivation of the EE and the charge fluctuations for graphene using its dimensional reduction to the one-dimensional SSH model. Our study shows that for models where the low-energy properties are described by Dirac fermions, the flux response of the EE is indeed exactly the one predicted from CFT. This response does not depend on the geometry of the entangling surface as long as it encloses the flux. We also considered the corner contributions to the EE. A standard way to extract universal quantities from the EE for gapped two-dimensional systems is via subtraction schemes. We showed that the usual subtraction schemes such as the Kitaev-Preskill cut, are not suitable for quantum critical points such as massless Dirac fermions. Whereas corner contributions are suppressed for gapped systems in such schemes, here they yield non-universal, geometry dependent results. However, we provide another subtraction scheme capable of cancelling out the area law and providing a direct access to the universal corner contributions.

More saliently, we proved that the flux dependence of the charge variance exhibits the same universal robustness. Despite its experimental relevance, this quantity has not been computed, to our knowledge, in the CFT framework. The dependence of the charge variance is exactly the same as that of the EE, up to a different constant prefactor. This work solely considered non-interacting fermions. It would be interesting to investigate if the features of charge fluctuations would convey to strongly interacting Dirac quantum system such as gapless Dirac quantum spin liquids. Being both simple to evaluate numerically and experimentally relevant, charge fluctuations could be an efficient probe for these systems. Another open question is whether the flux-dependence of the particle fluctuations requires particle conservation.

\acknowledgments

We thank Jean-Marie St\'ephan, Blagoje Oblak, Andrei Bernevig, Loic Herviou and William Witczak-Krempa for fruitful discussions. V.C., B.E. and N.R. were supported by Grant No. ANR-17-CE30-0013-01. NR was also partially supported by the DOE Grant No. DE-SC0016239, the Schmidt Fund for Innovative Research, Simons Investigator Grant No. 404513, and the Packard Foundation. Further support was provided by the NSF-EAGER No. DMR 1643312, NSF-MRSEC No. DMR-1420541 and DMR-2011750, ONR No. N00014-20-1-2303, Gordon and Betty Moore Foundation through Grant GBMF8685 towards the Princeton theory program, BSF Israel US foundation No. 2018226, and the Princeton Global Network Funds. A.H. acknowledges support by the European Research Council (ERC) under the European Union's Horizon 2020 research and innovation programme through the ERC Starting Grant WASCOSYS (No. 636201) and the ERC Consolidator Grant SEQUAM (No. 863476), and by the Deutsche Forschungsgemeinschaft (DFG) under Germany's Excellence Strategy (EXC-2111 – 390814868).

\bibliography{dirac.bib}

\appendix \newpage

\section{Correlation matrix} \label{app:CorrelationMatrix}

In this appendix, we consider translational invariant lattice models, and give efficient ways to evaluate their correlation matrix $C_{\mathcal{A}}$. 
Let us first express the lattice Hamiltonian as 
\begin{equation}
\mathcal{H} = \int_{BZ} \frac{{\rm d} \vec{k}}{V_{BZ}} \vec{c}^\dagger (\vec{k}) \tilde{h} (\vec{k}) \vec{c}(\vec{k}) \, ,
\end{equation}
in terms of Fourier transformed fermionic operators
\begin{equation}
c_\tau (\vec{r}) = \int_{BZ} \frac{{\rm d}\vec{k}}{ V_{BZ} } e^{ i \vec{k} \cdot \vec{r} } c_\tau (\vec{k}) \, ,
\end{equation}
and with $V_{BZ}$ the volume of the Brillouin zone $BZ$. 
Because $\vec{k}$ is a good quantum number, the correlation matrix is block-diagonal in momentum space $\Lambda_{\tau \tau'} (\vec{k}) = \Tr [\rho_T c_\tau^\dagger (\vec{k}) c_{\tau'} (\vec{k}) ]$. 
Its explicit expression
\begin{equation} \label{eqapp:Method_ExplicitCorrelMatrixGeneral}
\Lambda (\vec{k}) = \left[ 1 + e^{\beta \tilde{h} (\vec{k})} \right]^{-1} \, ,
\end{equation}
is straightforwardly derived from the Fermi-Dirac distribution of $\tilde{h} (\vec{k})$ eigenstates.

The real-space correlation matrix is obtained as
\begin{equation} \label{eqapp:Method_FFTCorrelMatrix} \begin{split}
C_{\alpha \beta}(\vec{r},\vec{r}') & = \Tr \left( \rho_T \, c_\alpha^\dagger (\vec{r}) c_\beta (\vec{r}') \right) \\
& = \int_{BZ} \frac{{\rm d} \vec{k}}{V_{BZ}} e^{-i \vec{k} \cdot (\vec{r}-\vec{r}')} \Lambda_{\alpha \beta} (\vec{k}) \, ,
\end{split} \end{equation}
and its restriction to $\vec{r},\vec{r}'$ in $\mathcal{A}$ yields $C_\mathcal{A}$. From the model-dependent $\tilde{h}$, Eq.~\ref{eqapp:Method_FFTCorrelMatrix} can either be evaluated analytically as in Sec.~\ref{ssec:SSHmodel} and Sec.~\ref{ssec:GrapheneExactLatticeCalculation}, or numerically with fast Fourier transform algorithms. In both cases, obtaining $C_\mathcal{A}$ is fast compared to its diagonalization. 

More generically, $\Lambda(\vec{k})$ can be obtained analytically for any two-band models ($d=2$). The hermitian Hamiltonian matrix can be written as a Pauli vector
\begin{equation}
h(\vec{k}) = d_0 (\vec{k}) + \vec{d} (\vec{k}) \cdot \vec{\sigma} \, ,
\end{equation}
with $\vec{\sigma} = (\sigma_x, \sigma_y, \sigma_z)$ the set of Pauli matrices. We can take its exponential, and find
\begin{equation} \label{eq:Method_ExplicitCorrelTwoBand}
\Lambda (\vec{k}) = \frac{1}{2} \left[ 1- u \frac{\vec{d}}{|\vec{d}|} \right] \, , \quad u = \frac{\cosh(\beta |\vec{d}|) - e^{-\beta d_0} }{  \sinh(\beta |\vec{d}|) } \,  .
\end{equation}
It is worth noting the particularly simple form $u=\tanh(\beta |\vec{d}|/2)$ when $d_0 = 0$, or $u=1$
if we furthermore work at zero temperature.

\section{Charge variance in the SSH chain}\label{app:SSHParticleVariance}

In this appendix we derive the expression Eq.~\ref{eq:SSH_Asymptotics_Variance} for the charge variance in the SSH chain. As mentioned in the main text the eigenvalues of the correlation matrix $C_\mathcal{A}$ in $]0,1[$ converge as $w \to \infty$ to   
 \begin{align}
\lambda_m = \frac{1}{1+q^{m}}  \quad \textrm{with } \left\{ \begin{array}{cc}  m \textrm{ odd } & \textrm{if } \delta <0   \\   m \textrm{ even } & \textrm{if } \delta >0   \end{array} \right. 
\end{align}  
with each value being doubly degenerate and where
\begin{equation}
q = e^{-\pi \frac{I(k') }{I(k)}}\label{eq:SSHqdefintion}
\end{equation}
The variance is given by 
 \begin{align}
V_{\mathcal{A}}^{\rm SSH} = 2 \sum_m  \lambda_m (1-\lambda_m) =    \sum_m \frac{1}{2 \cosh^2 \left( \frac{m \pi I(k')}{2I(k)} \right)}
\end{align} 
with $m$ is even or odd depending on the sign of $\delta$.
Using the $\textrm{sn}^2(z,k')$ Jacobi elliptic function, we get the following relation~\cite{Dunne_2000}
\begin{align}
k'^2 ~& \textrm{sn}^2(z,k')  = \frac{E(k)}{I(k)} \nonumber\\ 
  &-  \left(\frac{\pi}{I(k)} \right)^2 \sum_{m = -\infty}^{\infty} \frac{1}{4 \cosh^2  \left(\frac{\pi}{2 I(k)} \left( 2m I(k') - z \right)\right) }
\end{align}
where  $E(k)$ is the elliptic integral of the second kind
\begin{align}
E(k) = \int_0^{\frac{\pi}{2}} \sqrt{1+ k^2 \sin^2 \theta} d\theta
\end{align}
Taking $z=0$ (for $\delta>0$) and $z = I(k')$ (for $\delta<0$) yields Eq.~\ref{eq:SSH_Asymptotics_Variance}, namely 
\begin{align}
V_{\mathcal{A}}^{\rm SSH} = 2\frac{E(k) I(k)}{\pi^2} + 2 (k^2-1)\frac{I(k)^2}{\pi^2}, \qquad (\delta <0)
\end{align}
and
\begin{align}
V_{\mathcal{A}}^{\rm SSH} = 2\frac{E(k) I(k)}{\pi^2}, \qquad (\delta >0)\,.
\end{align}
In both regimes the variance diverges as $\delta \to 0$ as 
\begin{align}
V_{\mathcal{A}}^{\rm SSH}  \sim  \frac{1}{\pi^2} \log \xi_{\rm SSH} \sim  - \frac{1}{\pi^2} \log |\delta| 
\end{align}
and therefore 
\begin{align}
S_{\mathcal{A}}^{\rm SSH} \sim \frac{\pi^2}{3}V_{\mathcal{A}}^{\rm SSH}, \qquad (\delta \to 0) \,.
\end{align}
Such a behavior is expected as soon as charge fluctuations become gaussian, in the sense that the higher cumulants are suppressed relatively to the charge variance~\cite{Klich_2006,Klich_2009a,Calabrese_2012}. This is indeed what happens in the SSH chain when the correlation length $\xi_{\rm SSH}$ becomes large, \emph{i.e.} in the critical regime. To see this, we can exploit the fact that the full counting statistics is known exactly for the SSH chain, via the cumulant generating function
\begin{align}
f_\mathcal{A}(t) = \log \langle e^{t N_A} \rangle
\end{align}
This generating function has been evaluated in Ref.~\cite{jin2007entropy}, yielding:
\begin{align} 
f_\mathcal{A}(t) =  t  w + 4\log \frac{\theta_j \left(  \frac{t }{2\pi i}  | \tau \right)}{\theta_j \left(0 | \tau \right)} + O(w^{-\infty})
 \end{align}
where $\tau = i I(k')/I(k)$, $j=2$ for $\delta <0$, and $j=3$ for $\delta >0$. In order to analyse the behavior close to criticality ($\delta \to 0$, thus $\tau \to 0$), it is more convenient to write  (using the modular properties of theta functions)
    \begin{align}
f_\mathcal{A}(t) & =  t  w+  \frac{1}{- i \tau} \frac{t^2}{\pi} + 4 \log  \frac{\theta_j \left( \frac{t}{2\pi i \tau}    |  -\frac{1}{\tau} \right)}{\theta_j \left(0 | -\frac{1}{\tau} \right)}  + O(w^{-\infty})
\end{align}
with $j=3$ for $\delta <0$ and $j=4$ for $\delta >0$. From the above expression it appears that only the term in $t^2$, that is the charge variance, blows up as $\delta \to 0$, while the other (even) cumulants remain finite. Note that the odd cumulants vanish identically, as expected for a semi-infinite interval, due to the relation $\kappa_n(A) = (-1)^n \kappa_n(B)$ for the $n^{\rm th}$ cumulant.

\section{Free fermion models with Dirac modes in 2d} \label{app:TwoDimFerionicModels}

In this appendix, we review the definitions and the main properties of the two dimensional tight-binding Hamiltonians hosting Dirac cones used in Sec.~\ref{ssec:OtherModels}. Their Bravais lattice, Bloch Hamiltonian, parameters and their number of Dirac cones are summarized in Tab.~\ref{apptab:ModelTwoDim}.

The three first lines of Tab.~\ref{apptab:ModelTwoDim} describe model with a hexagonal Bravais lattice. We use the conventions and notations introduced in Sec.~\ref{ssec:GrapheneExactLatticeCalculation}. The first line represents the tight-binding model of graphene studied in the main text (see Sec.~\ref{ssec:GrapheneExactLatticeCalculation}). It has two Dirac cones at the $K$ and $K'$ points of the $BZ$. Carefully introducing and tuning next-nearest neighbor hopping and staggered potential, it is possible to open a gap at $K'$ while keeping a Dirac cone at $K$. This corresponds to the Haldane model on the critical line~\cite{haldane1988model}, which appears on the second line of Tab.~\ref{apptab:ModelTwoDim}. The third line depicts nearest neighbor hopping model on the Kagome lattice, where we have added an energy shift equal to the tunneling amplitude in order to bring the two Dirac cones (also at the $K$ and $K'$ points) to zero energy. 

The fourth and fifth lines of Tab.~\ref{apptab:ModelTwoDim} show models defined on a square Bravais lattice, each having two orbitals per unit cell. We choose the following basis vectors 
\begin{equation}
\vec{a}_1 = \left( 1,0 \right) \, , \quad \vec{a}_2 = \left( 0,1 \right) \, ,
\end{equation}
and the periodic boundary conditions along $x$ and $y$ allows to identify any point of the lattice $\vec{r}$ with both $\vec{r}+N_x \vec{a}_1$ and $\vec{r}+N_y\vec{a}_2$. We use the first $BZ$ associated with this lattice, \textit{i.e.} $\vec{k}=(k_x,k_y)$ with $k_x,k_y \in (-\pi,\pi]$. Only the mass $M$ differs between the \sfrac{1}{2}-BHZ~\cite{bernevig2006quantum} and QWZ models~\cite{qi2006topological}, but it changes both the number and position of the Dirac cones in the problem, as described in Tab.~\ref{apptab:ModelTwoDim}. 

Lastly, we consider the $\pi$-flux model. It is defined on the square lattice and has two orbitals per unit cell labeled $\tau = A$ and $B$. Tunneling amplitudes are equal in magnitude but their signs differ and define as below: 
\begin{itemize}
\item Along horizontal links, all nearest neighbor $A-B$ links have a positive tunneling amplitudes. 
\item Along vertical links, nearest neighbor $A-A$ (resp. $B-B$) links have positive (resp. negative) tunneling coefficients. 
\item There is no tunneling on horizontal $A-A$ and $B-B$ links, nor on vertical $A-B$ ones. 
\end{itemize}
This pattern leads to the Bloch Hamiltonian given in the last line of Tab.~\ref{apptab:ModelTwoDim}.

\begin{table*} 
\centering \label{apptab:ModelTwoDim}
\caption{List of two dimensional tight-binding models studied in the main text in Sec.~\ref{ssec:OtherModels}. The first column is the model name, the second column is the Bravais lattice. The third column gives the Bloch Hamiltonian. The fourth column provides the number and location of the Dirac points. The last column gives additional information about the Bloch Hamiltonian parameters.}
\begin{tabular}{| c || c | c | c | C{0.4\textwidth} |} 
\hline
Name & Lattice & Bloch Hamiltonian & Dirac cones & Additional information \\
\hline\hline
Graphene & Honeycomb & $\begin{bmatrix} 0 & f^* \\ f & 0 \end{bmatrix}$ & $\frac{2\pi}{3\sqrt{3}} \left( \pm \sqrt{3} , 1 \right)$  & $f = 1 + e^{i\vec{k}\cdot\vec{a}_1} + e^{i\vec{k}\cdot\vec{a}_2}$, see text  \\
\hline
Haldane & Honeycomb & $\begin{bmatrix} g & f^* \\ f & -g \end{bmatrix}$ & $\frac{2\pi}{3\sqrt{3}} \left( - \sqrt{3} , 1 \right)$  & On the critical line: $g = 3\sqrt{3} + 2 [\sin (\vec{k}\cdot \vec{a}_1) - \sin (\vec{k}\cdot \vec{a}_2) + \sin (\vec{k}\cdot (\vec{a}_2-\vec{a}_1))]$ \\
\hline
Kagome & Kagome & $ 1 - \begin{bmatrix} 0&c_2&c_3\\c_2&0&c_1\\c_3&c_1&0 \end{bmatrix}$  & $\frac{2\pi}{3\sqrt{3}} \left( \pm \sqrt{3} , 1 \right)$ & The energy shift $1$ brings the two Dirac cones at zero energy \\
\hline
\sfrac{1}{2}-BHZ & Square & $\begin{bmatrix} M-c_x-c_y & s_x - i s_y \\ s_x + i s_y & -M+c_x+c_y \end{bmatrix}$ & $(0,0)$  & $c_{x/y} = \cos(k_{x/y})$, $s_{x/y} = \sin(k_{x/y})$ and $M=2$~\cite{bernevig2006quantum} \\
\hline
QWZ & Square & $\begin{bmatrix} M-c_x-c_y & s_x - i s_y \\ s_x + i s_y & -M+c_x+c_y \end{bmatrix}$ & $(0,\pi)$, $(\pi,0)$  & $c_{x/y} = \cos(k_{x/y})$, $s_{x/y} = \sin(k_{x/y})$ and $M=0$~\cite{qi2006topological} \\
\hline
$\pi$-flux & Square & $\begin{bmatrix} c_y & c_x \\ c_x & - c_y \end{bmatrix}$ & $\left( \pm \frac{\pi}{2}, \pm \frac{\pi}{2} \right)$ & --- \\
\hline
\end{tabular}
\end{table*}

\section{Three-dimensional chiral hinge model} \label{app:HingeModel}

\begin{figure}
\centering
\includegraphics[width=0.65 \columnwidth]{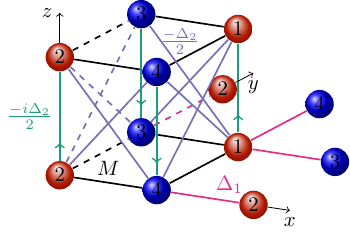}%
\caption{Tight-binding model for the three-dimensional chiral hinge model. The model is defined on a cubic lattice with a unit cell of four sites lying in the $(x,y)$ plane, labeled $\tau=1, 2, 3$ and $4$. In this plane, sites in the same unit cell are connected by a nearest-neighbour hopping $M$ marked by black lines ($-M$ for dashed black lines). In the $(x,y)$ plane, sites in adjacent unit cells are connected by a nearest-neighbour hopping $\Delta_1$ marked by violet lines ($-\Delta_1$ for dashed violet lines). In the $z$ direction, adjacent unit cells are connected by a real next-nearest neighbour hopping $- \Delta_2/ 2$ marked by light blue lines ($\Delta_2/ 2$ for dashed light blue lines). In addition, there is a purely imaginary nearest neighbour hopping between adjacent unit cells in the $z$ direction with value $-i\Delta_2 / 2$ in the direction of the green arrows. We study the model for parameter values $M = \Delta_1 = \Delta_2 = 1$.} 
\label{fig:SketchCHI}
\end{figure}

\begin{figure}
\centering
\includegraphics[width=0.99 \columnwidth]{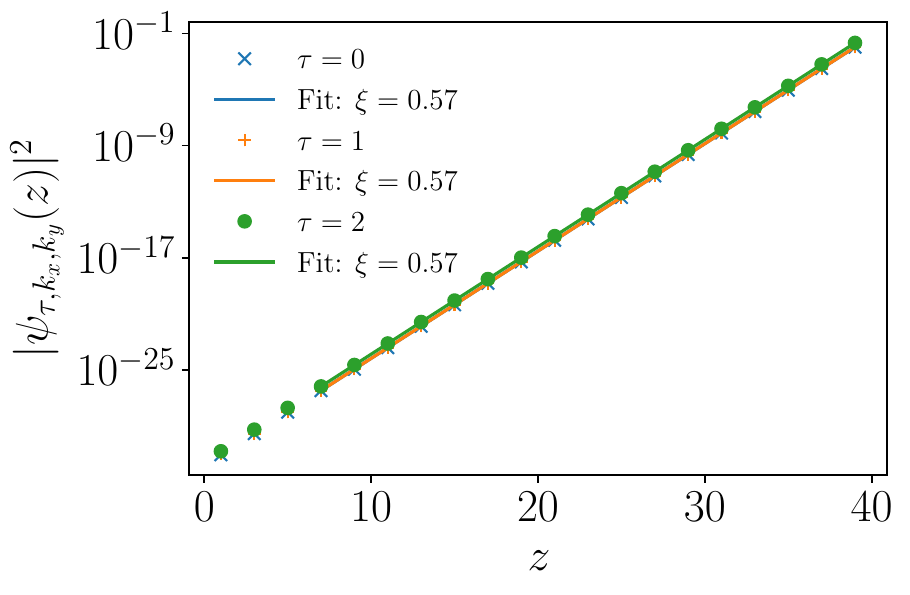}%
\caption{Exponential localization of the Dirac cones at the horizontal surfaces of the 3D chiral hinge insulator with $40 \times 10 \times 40$ unit cells. Shown is the weight $\lvert \psi_{\tau, k_x, k_y} (z)\lvert ^2$ of one out of the four single particle modes at surface momentum $(k_x, k_y) = K$ and zero energy as a function of the depth $z$ in the 3D bulk, resolved according to the four sublattices $\tau = 1, \dotsc, 4$. We picked a linear superposition such that the weight on the site $\tau = 2$ vanishes at the top surface $z = 0$. Due to symmetry, this yields two states whose weight is zero for all even values of $z$. From these two, we chose a linear superposition such that the weight on the site $\tau = 4$ vanishes at the bottom surface $z = 39$, which results in a state whose weight is zero on all sites with $\tau = 4$. The weight of the remaining three sublattices decays exponentially with a correlation length $\xi = 0.57$.} 
\label{fig:LocalisationSurfaceDD}
\end{figure}

In Sec.~\ref{ssec:OtherModels}, we have also considered the three-dimensional chiral hinge model of Ref.~\cite{schindler2018higher} defined on a cubic lattice with four sites in the $(x,y)$ plane per unit cell. We denote these four sites $\tau=1, 2, 3$ and $4$ (see Fig.~\ref{fig:SketchCHI}). At half filling, it realizes a second-order topological insulator~\cite{benalcazar2017quantizedScience}. With open boundaries in the $x$ and $y$ directions, the vertical surfaces are gapped, but the hinges parallel to the $z$ direction host one-dimensional gapless chiral modes. Moreover, the system hosts a single massless Dirac mode on each horizontal surface, that is on its top and bottom surfaces~\cite{schindler2018higher}, which are located at the momentum $K = (\pi, \pi)$ in the surface Brillouin zone. The Dirac cones are exponentially localized at the surfaces, as shown in Fig.~\ref{fig:LocalisationSurfaceDD}. We want to study if the EE and charge variance for the Dirac cone on one of these surfaces, say the top surface, satisfies the same flux dependence as a strictly two-dimensional system,
Eqs.~\ref{eq_sin_flux} and~\ref{eq:FluxDependenceOfVariance}.

\begin{figure}
\centering
\includegraphics[width=0.6\columnwidth]{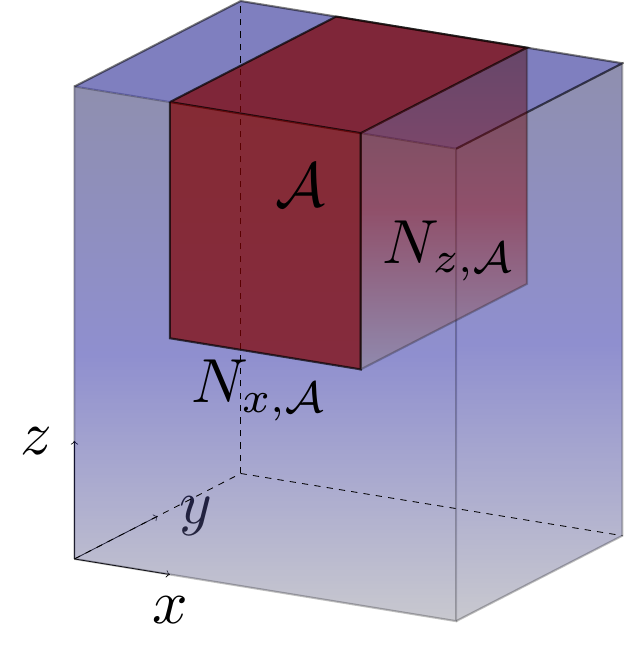}
\caption{Sketch of the geometry used for the EE computation in the 3D chiral hinge insulator with PBC in the $x$ and $y$ directions. The subsystem $\mathcal{A}$ includes a part of the top surface of width $N_{x,\mathcal{A}}$ in the $x$ direction, preserves translational symmetry in the $y$ direction and extends to a depth $N_{z,\mathcal{A}}$ into the three-dimensional bulk.} 
\label{fig:EE3DCHI}
\end{figure}

To that end, we consider the geometry sketched in Fig.~\ref{fig:EE3DCHI} with periodic boundary conditions in the $x$ and $y$ directions. The subsystem $\mathcal{A}$ includes a part of the top surface of width $N_{x,\mathcal{A}}$ in the $x$ direction, preserves translational symmetry in the $y$ direction and extends to a depth $N_{z,\mathcal{A}}$ into the three-dimensional bulk. We are interested in the dependence of the entropy $S_{\mathcal{A}}$ on the twist angle $\phi \in [0, 2\pi)$ of the boundary conditions in the $y$ direction. As in the two-dimensional case, the difference $S_{\mathcal{A}}(\phi) - S_{\mathcal{A}}(\pi)$ cancels all area law contributions originating from the two surfaces of $\mathcal{A}$ normal to the $x$ direction and the bottom surface of $\mathcal{A}$. Moreover, any potential hinge or corner contributions are also cancelled out. As shown in Fig.~\ref{fig:AllModelCollapse} of the main text, $S_{\mathcal{A}}(\phi) - S_{\mathcal{A}}(\pi)$ obeys the same scaling as in the two-dimensional case for open boundaries in the $z$ direction, provided that the relevant correlation length $N_y /\phi$ is small compared to $N_{x,\mathcal{A}}$ and $N_{z,\mathcal{A}}$. We have confirmed that this characteristic scaling is due entirely to the surface Dirac mode. Indeed, with periodic boundary conditions in the $z$ direction, for which no surface Dirac cone is present, the variation in $S_{\mathcal{A}}(\phi) - S_{\mathcal{A}}(\pi)$ is less than $1\%$ of the open boundary result for the same system and subsystem sizes.

\begin{figure}
\centering
\includegraphics[width=0.99\columnwidth]{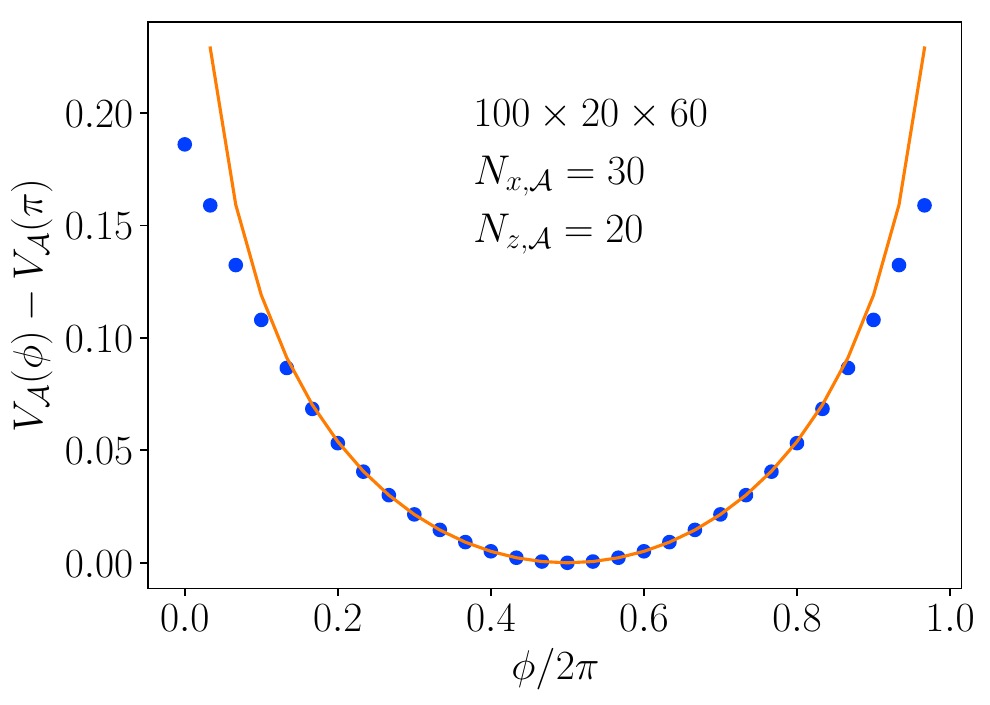}
\caption{Charge variance measured with respect to its value at $\phi=\pi$ for the surface Dirac cone of the chiral hinge insulator model. We use the geometry shown in Fig.~\ref{fig:EE3DCHI} with a total system size $(N_x ,N_y , N_z) = (100,20,60)$ and an entangling region of $(N_{x,\mathcal{A}} ,N_{y,\mathcal{A}} , N_{z,\mathcal{A}}) = (30,20,20)$. The solid orange line is the asymptotic prediction of Eq.~\ref{eq:FluxDependenceOfVariance}.} 
\label{fig:VarNACHI}
\end{figure}

We now consider the charge fluctuations and their flux dependence for this model like we did in Sec.~\ref{sec:NumericalChargeFluctuations} for the two-dimensional $\sfrac{1}{2}$-BHZ model. For that purpose we use the same entangling region $\mathcal{A}$ than previously and shown in Fig.~\ref{fig:EE3DCHI}. We use the same system and subsystem size as in Sec.~\ref{ssec:OtherModels} for the EE, namely $(N_x ,N_y , N_z) = (100,20,60)$ and $(N_{x,\mathcal{A}} ,N_{y,\mathcal{A}} , N_{z,\mathcal{A}}) = (30,20,20)$. Like for the EE, the contributions coming for the parts of $\mathcal{A}$ located in the bulk of the system are cancelled out by the subtraction of the variance at $\phi=\pi$. As shown in Fig.~\ref{fig:VarNACHI}, we once again see good agreement with the asymptotic expression of Eq.~\ref{eq:FluxDependenceOfVariance}. Finally, we stress that the current results and techniques for this chiral hinge insulator hold true for other Dirac states at the surface of insulators such as time-reversal invariant three-dimensional topological insulators.

\end{document}